\documentclass[review]{elsarticle}

\usepackage{lineno}
\modulolinenumbers[5]
\usepackage{amsmath}
\usepackage{amssymb}
\usepackage{amsthm}
\usepackage{booktabs}
\usepackage{color}
\usepackage{float}
\usepackage[T1]{fontenc}
\usepackage{graphicx}
\usepackage{subcaption}
\usepackage[bookmarks=false]{hyperref}
\usepackage{ifpdf}
\usepackage[utf8]{inputenc}
\usepackage{keyval}
\usepackage{listings}
\usepackage{makecell}
\usepackage{moresize}
\usepackage{multirow}
\usepackage{nicefrac}
\usepackage{paralist}
\usepackage{rotating}
\usepackage{setspace}
\usepackage{tabularx}
\usepackage{url}
\usepackage{wrapfig}
\usepackage{xcolor}
\usepackage{xspace}

\theoremstyle{definition}

\definecolor{listinggray}{gray}{0.95}
\definecolor{darkgray}{gray}{0.7}
\definecolor{commentgreen}{rgb}{0, 0.4, 0}
\definecolor{darkblue}{rgb}{0, 0, 0.4}
\definecolor{middleblue}{rgb}{0, 0, 0.7}
\definecolor{darkred}{rgb}{0.4, 0, 0}
\definecolor{brown}{rgb}{0.5, 0.5, 0}

\usepackage[normalem]{ulem}
\def\cyanuwave{\bgroup \markoverwith{\lower3.5\p@\hbox{\sixly \textcolor{cyan}{\char58}}}\ULon}
\def\reduwave{\bgroup \markoverwith{\lower3.5\p@\hbox{\sixly \textcolor{red}{\char58}}}\ULon}
\def\blueuwave{\bgroup \markoverwith{\lower3.5\p@\hbox{\sixly \textcolor{blue}{\char58}}}\ULon}
\font\sixly=lasy6 
\makeatother

\newif\ifdraft
\drafttrue
\ifdraft

\newcommand{\jhanote}[1]{ {\textcolor{red} { ***shantenu: #1 }}}
\newcommand{\gpnote}[1]{{\textcolor{green} {***giannis: #1}}}
\newcommand{\mtnote}[1]{{\textcolor{orange} {***matteo: #1}}}
\newcommand{\aanote}[1]{{\textcolor{blue} {***aymen: #1}}}
\newcommand{\todo}[1]{ {\textcolor{brown} { TODO #1 }}}
\newcommand{\note}[1]{ {\textcolor{magenta} { ***Note: #1 }}}
\else
\newcommand{\jhanote}[1]{}
\newcommand{\gpnote}[1]{}
\newcommand{\mtnote}[1]{}
\newcommand{\todo}[1]{}
\newcommand{\note}[1]{}
\newcommand{\aanote}[1]{}
\fi

\newcommand{\entk}{EnTK\xspace}

\floatstyle{ruled}
\newfloat{scheme}{htbp}{lok}
\floatname{scheme}{Scheme}

\lstdefinestyle{myListing}{
    frame=single,
    backgroundcolor=\color{listinggray},
    language=C,
    basicstyle=\ttfamily \footnotesize,
    breakautoindent=true,
    breaklines=true
    tabsize=2,
    captionpos=b,
    aboveskip=0em,
    belowskip=-2em,
}

\lstdefinestyle{myPythonListing}{
    frame=single,
    backgroundcolor=\color{listinggray},
    language=Python,
    basicstyle=\ttfamily \scriptsize,
    breakautoindent=true,
    breaklines=true
    tabsize=2,
    captionpos=b,
}

\ifpdf
\DeclareGraphicsExtensions{.pdf, .jpg, .tif}
\else
\DeclareGraphicsExtensions{.ps,  .eps, .jpg}
\fi

\usepackage{listings}
\usepackage{paralist}
\lstnewenvironment{code}[1][]%
{
    \noindent
    \minipage{1.0 \linewidth}
    \vspace{0.5\baselineskip}
    \lstset{
        language=Python,
        frame=single,
        captionpos=b,
        stringstyle=\ttfamily,
        basicstyle=\scriptsize\ttfamily,
        showstringspaces=false,#1}
}
{\endminipage}
\begin{document}
    
\begin{frontmatter}

	\title{Comparing Workflow Application Designs for High Resolution Satellite Image Analysis}
	
	\author[mainaddress]{Aymen Al-Saadi}
	\author[mainaddress]{Ioannis Paraskevakos}
	\author[secondaddress]{Bento Collares Gonçalves}
	\author[secondaddress]{Heather J. Lynch}
	\author[mainaddress, thirdaddress]{Shantenu Jha}
	
	\author[mainaddress]{Matteo Turilli\corref{mycorrespondingauthor}}
	\cortext[mycorrespondingauthor]{Corresponding author}
	
	\address[mainaddress]{Department of Electrical and Computer Engineering, Rutgers University, Piscataway, New Jersey 08854}
	\address[secondaddress]{Department of Ecology and Evolution, Stony Brook, NY USA 11777}
	\address[thirdaddress]{Brookhaven National Laboratory}
	
\begin{abstract}
    Very High Resolution satellite and aerial imagery are used to monitor and
    conduct large scale surveys of ecological systems. Convolutional Neural
    Networks have successfully been employed to analyze such imagery to detect
    large animals and salient features. As the datasets increase in volume and
    number of images, utilizing High Performance Computing resources becomes
    necessary. In this paper, we investigate three task-parallel, data-driven
    workflow designs to support imagery analysis pipelines with heterogeneous
    tasks on HPC\@. We analyze the capabilities of each design when processing
    datasets from two use cases for a total of 4,672 satellite and aerial
    images, and 8.35~TB of data. We experimentally model the execution time of
    the tasks of the image processing pipelines. We perform experiments to
    characterize the resource utilization, total time to completion, and
    overheads of each design. Based on the model, overhead and utilization
    analysis, we show which design is best suited to scientific pipelines with
    similar characteristics.
\end{abstract}

	\end{frontmatter}
	

\section{Introduction}\label{sec:intro}

A growing number of scientific domains are adopting workflows that use
multiple analysis algorithms to process a large number of images. The volume
and scale of data processing justifies the use of parallelism, tailored
programming models and high performance computing (HPC) resources. While
these features create a large design space, the lack of architectural and
performance analyses makes it difficult to chose among functionally
equivalent implementations.

In this paper we focus on the design of computing frameworks that support the
execution of heterogeneous tasks on HPC resources to process large imagery
datasets. These tasks may require one or more CPUs and GPUs, implement diverse
functionalities and execute for different amounts of time. Typically, tasks
have data dependences and are therefore organized into workflows. Due to task
heterogeneity, executing workflows poses challenges of effective scheduling,
correct resource binding and efficient data management. HPC infrastructures
exacerbate these challenges by privileging the execution of single,
long-running tasks.

From a design perspective, a promising approach to address these challenges
is isolating tasks from execution management. Tasks are assumed to be
self-contained programs which are executed in the operating system (OS)
environment of HPC compute nodes. Programs implement the domain-specific
functionalities required by use cases while computing frameworks implement
resource acquisition, task scheduling, resource binding, and data management.

Compared to approaches in which tasks are functions or methods, a
program-based approach offers several benefits as, for example, simplified
implementation of execution management, support of general purpose programming
models, and separate programming of management and domain-specific
functionalities. Nonetheless, program-based designs impose performance
limitations, including OS-mediated intertask communication and task spawning
overheads, as programs execute as OS processes and do not share a memory
space.

Due to their performance limitations, program-based designs of computing
frameworks are best suited to execute compute-intensive workflows in which
each task requires a certain amount of parallelism and runs from several
minutes to hours. The use of modern HPC infrastructures with large numbers of
CPUs/GPUs presents new challenges to the design of program-based workflows
that require heterogeneous, compute-intensive tasks that process large
amounts of data.

We use two paradigmatic use cases from the polar science domain to evaluate
three alternative designs of computing frameworks for executing program-based
tasks, and experimentally characterize and compare their performance. The
first use case requires us to detect pack-ice seals by analyzing satellite
images of Antarctica taken across a whole calendar year. The resulting dataset
consists of 3,097 images for a total of 4~TB\@. This use case requires us to
repeatedly process these images, running tasks on both CPUs and GPUs  that
exchange several GB of data. The second use case requires us to match paired
images of penguin colonies from Antarctica and estimate the approximate
location where images were taken. The dataset contains 1,575 images for a
total of 1~TB\@.

The first design uses a pipeline to independently process each image, while
the second and third designs use the same pipeline to process a series of
images with differences in how images are bound to available compute nodes.

This paper offers four main contributions: 
\begin{inparaenum}[(1)]
    \item a GPU implementation of the Scale Invariant Fast Transformation
    (SIFT) algorithm to serve the purpose of geolocating satellite imagery
    specifically;
    \item an indication of how to further the implementation of our workflow
    engine so as to support the class of use cases we considered, while
    minimizing workflow time to completion and maximizing resource
    utilization;
    \item specific design guidelines for supporting data-driven,
    compute-intensive workflows on high-performance computing resources with
    a task-based computing framework; and
    \item an experiment-based methodology to compare performance of
    alternative designs that does not depend on the use case and computing
    framework presented in this paper.
\end{inparaenum}

The paper is organized as follows. \S\ref{sec:related} provides a survey of
the state of the art. \S\ref{sec:ucase} presents the use cases in more detail
and discusses their computational requirements as well as the individual
stages of the pipelines. \S~\ref{sec:gpu_imp} describe and discuss the novel
implementation of SIFT and its performance. \S\ref{sec:w-design} discusses
the three program-based designs in detail. \S\ref{sec:experiments} details
our performance evaluation, discussing the results of our experiments.
In~\S\ref{sec:conclusion}, we summarize the contributions of this paper and
identify to some new lines of research that it opens.

\section{Related Work}\label{sec:related}

Several tools and frameworks are available for image analysis based on
diverse designs and programming paradigms, and implemented for specific
resources. Numerous image analytics frameworks for medical, astronomical, and
other domain specific imagery provide MapReduce~\cite{dean2010mapreduce}
implementations. MaReIA~\cite{vo2018mareia}, built for medical image
analysis, is based on Hadoop and Spark~\cite{zaharia2010spark}.
Kira~\cite{zhang2016kira}, built for astronomical image analysis, also uses
Spark and pySpark, allowing users to define custom analysis applications.
Further, Ref.~\cite{yan2014large} proposes a Hadoop-based cloud Platform as a
Service, utilizing Hadoop's streaming capabilities to reduce filesystem reads
and writes. These frameworks support clouds and/or commodity clusters for
execution.

BIGS~\cite{ramos2012bigs} is a framework for image processing and analysis.
BIGS is based on the master-worker model and supports heterogeneous
resources, such as clouds, grids and clusters. BIGS deploys a number of
workers to resources, which query its scheduler for jobs. When a worker can
satisfy the data dependencies of a job, it becomes responsible to execute it.
BIGS workers can be deployed on any type of supported resource. The user is
responsible for defining the input, processing pipeline, and launching BIGS
workers. As soon as a worker is available, execution starts. In addition,
BIGS offers a diverse set of APIs for developers. BIGS approach is very close
to Design 1 we described in~\S\ref{des1}.

LandLab~\cite{hobley2017creative} is a framework for building, coupling and
exploring two-dimensional numerical models for Earth-surface dynamics.
LandLab provides a library of processing constructs. Each construct is a
numerical representation of a geological process. Multiple components are
used together, allowing the simulation of multiple processes acting on a
grid. The design of each component is intended to work in a plug-and-play
fashion. Components couple simply and quickly. Parallelizing Landlab
components is left to the developer.

The High Performance Computing (HPC)/ High Throughput Computing (HTC)
Software Infrastructure for the Synthesis and Analysis of Cosmic Microwave
Background (CMB) Datasets~\cite{Borrill2020} is a project to enable CMB
experiments to seamlessly use both HPC and HTC systems for their simulation
and processing needs. Specifically, this project develops compatible data
models to enable bidirectional data flow among pipeline components
concurrently executing on HPC, HTC and hybrid infrastructures. The project
extends the Time Ordered Astrophysics Scalable Tools (TOAST) to support data
translation and unification across these infrastructures.

The Sea Ice High Resolution Image Analytics (ArcCI)~\cite{yan2014large} is a
framework that uses cloud computing for big data management and
visualization. ArcCI is implemented as a set of web services to collect,
search, explore, visualize, organize, analyze and share collections of high
spatial resolution Arctic sea ice imagery. Currently, ArcCI supports 35
datasets for a total of 1.96~TB of data.

The Large-scale IMage Processing Infrastructure Development
(LIMPID)~\cite{Manjunath2018} project developed the Bio-Image Semantic Query
User Environment (BisQue) for managing, analyzing and sharing images and
metadata for large-scale problems. The main goal of BisQue is to enable
reproducible image data science on cloud platforms, supporting multiple
imaging modalities such as photographs, satellites and microscopes.

Image analysis libraries, frameworks and applications have been proposed for
HPC resources. PIMA(GE)\textsuperscript{2} Library~\cite{galizia2015mpicuda}
offers a low-level API for parallel image processing using MPI and CUDA\@.
SIBIA~\cite{gholami2017framework} is a framework for coupling biophysical
models with medical image analysis, providing users parallel computational
kernels through MPI and vectorization. Ref.~\cite{huo2018towards} proposes a
scalable medical image analysis service. This service uses
DAX~\cite{damon2017dax} as an engine to create and execute image analysis
pipelines. Tomosaic~\cite{vescovi2018tomosaic} is a Python framework, used
for medical imaging, employing MPI4py to parallelize different parts of the
workflow.

Petruzza et al.~\cite{petruzza2017isavs} describe a scalable image analysis
library. Their approach defines pipelines as data-flow graphs, with user
defined functions as tasks. Charm++ is used as the workflow management layer,
by abstracting the execution level details, allowing execution on local
workstations and HPC resources. Teodoro et
al.~\cite{teodoro2013highthroughput} define a master-worker framework
supporting image analysis pipelines on heterogeneous resources. The user
defines an abstract dataflow and the framework is responsible for scheduling
tasks on CPU or GPUs. Data communication and coordination is done via MPI\@.
Ref.~\cite{grunzke2017seamless} proposes the use of
UNICORE~\cite{benedyczak2016unicore} to define image analysis workflows on
HPCs.

Image classification is an existing problem of interest for computer vision
scientists. The most common approaches are using scene and object recognition
technology. These approaches identify a set of images and classify them based
on the scene of interest, e.g., building, mountain, or lakes. A disadvantage
of these approaches is that they do not estimate the approximate geographical
location of images.

Another approach in computer vision is geolocating satellite and ground-level
imagery. Ghouaiel and Lefèvre~\cite{Ghouaiel2016CouplingGP} proposed an
automatic translation for ground photos into aerial viewpoint, the technique
specifically supports only wide panoramic photos with an accuracy of 54\%.

In the large scale image geolocalization, the approach is based on using the
``IM2GPS'' algorithm~\cite{vo2017revisiting}. IM2GPS uses a convolutional
neural network (CNN) to geolocalize images against a database of geotagged
Internet photographs, used as training data. The approach reaches an accuracy
of 25\% for the 237 photos in their dataset.

We introduce another image geolocating approach based on the image matching
technique to extract the similarity level between two images and estimate the
approximate location as values of longitude and latitude. We focus on
geolocating a set of historic aerial photo imagery using satellite imagery as
a basemap.

Our workflow approach proposes designs for image analysis pipelines that are
domain independent, i.e., not specific to medical, astronomical, or other
domain imagery. Both the workflow and runtime systems we use allow execution
on multiple HPC resources with no change in our approach, independent from
the types, durations and sizes of task that workflows require to execute.
Furthermore, in one of the proposed designs parallelization is inferred,
allowing correct execution regardless of the multi-core or multi-GPU
capabilities of the used resource.

All the above, except Ref.~\cite{zhang2016kira}, focus on characterizing the
performance of the proposed solution. Ref.~\cite{zhang2016kira} compares
different implementations, one with Spark, one with pySpark, and an MPI
C-based implementation. This comparison is based on the weak and strong
scaling properties of the approaches. Our approach offers a well-defined
methodology to compare different designs for task-based and data-driven
pipelines with heterogeneous tasks.

\section{Satellite Imagery Analysis Use Cases}\label{sec:ucase}

In this paper we developed and characterized computational workflows that
satisfy the requirements of two earth science use cases. The first use case,
labeled as UC1, requires to process imagery to find Antarctic pack-ice seals.
The second use case, labeled as UC2, geolocates an aerial image using a
satellite image as a basemap. These use cases require to develop application
workflows in which images are processed and analyzed in multiple stages in
order to find some relevant properties. This application pattern is used in
many scientific domains and, as such, our two use cases are paradigmatic of a
common set of computing requirements.

\subsection{Seals Use Case (UC1)}\label{ssec:seals-uc}

The imagery employed by ecologists as a tool to survey populations and
ecosystems come from a wide range of sensors, e.g., camera-trap
surveys~\cite{karanth1995estimating} and aerial imagery
transects~\cite{western2009impact}. However, most traditional methods can be
prohibitively labor-intensive when employed at large scales or in remote
regions. Very High Resolution (VHR) satellite imagery provides an effective
alternative to perform large scale surveys at locations with poor
accessibility such as surveying Antarctic fauna~\cite{lynch2012detection}. To
take full advantage of increasingly large VHR imagery, and reach the spatial
and temporal breadths required to answer ecological questions, it is
paramount to automate image processing and labeling.

Convolutional Neural Networks (CNN) represent the state-of-the-art for nearly
every computer vision routine. For instance, ecologists have successfully
employed CNNs to detect large mammals in airborne
imagery~\cite{kellenberger2018detecting,polzounov2016right} and camera-trap
survey imagery~\cite{norouzzadeh2018automatically}. We use a CNN to survey
Antarctic pack-ice seals in VHR imagery. Pack-ice seals are a main component
of the Antarctic food web~\cite{fabra2008convention}; estimating the size and
trends of their populations is key to understanding how the Southern Ocean
ecosystem copes with climate change~\cite{hillebrand2018climate} and
fisheries~\cite{reid2019climate}.

For this use case, we process WorldView 3 (WV03) panchromatic imagery as
provided by DigitalGlobe Inc. This dataset has the highest available
resolution for commercial satellite imagery. We refrain from using imagery
from other sensors because pack-ice seals are not clearly visible at lower
resolutions. For our CNN architecture, we use a U-Net~\cite{ronneberger2015u}
variant that counts seals with an added regression branch and locates them
using a seal intensity heat map. To train our CNN, we use a training set of
53 WV03 images, with 88,000 hand-labeled tiles, where every tile has a
correspondent seal count and a class label (i.e., seal vs. non-seal). For
hyper-parameter search, we train CNN variants for 75 epochs (i.e., 75
complete runs through the training set) using an Adam
optimizer~\cite{kingma2014adam} with a learning rate of $10^{-3}$ and tested
against a validation set. The validation set consists of 10\% of the
training set. In addition, we randomized the WV03 image selection so that a
validation tile does not overlap with a training one. Testing was performed
on 5 WV03 images. Double observer seal counting was performed and model
detection results were compared to observer consensus detections.
Furthermore, we avoided double counting by setting a minimum distance
boundary between neighboring seals and keeping those where the model was more
confident.

We use the best performing model on an archive of over 3,097 WV03 images,
with a total dataset size of 4~TB\@. Due to limitations on GPU memory, it is
necessary to tile WV03 images into smaller patches before sending input
imagery through the seal detection CNN\@. Taking tiled imagery as input, the
CNN outputs the latitude and longitude of each detected seal. While the raw
model output still requires statistical treatment, such ``mock-run'' emulates
the scale necessary to perform a comprehensive pack-ice seal census. We order
the tiling and seal detection stages into a pipeline that can be re-run
whenever new imagery is obtained. This allows domain scientists to create
seal abundance time series that can aid in Southern Ocean monitoring.

\subsection{Image Geolocation Use Case (UC2)}\label{ssec:geoloc-uc}

Image geolocating or geotagging is the process of appending geographical
identification metadata to images. Each image is
paired to other images and specialized algorithms are used to extract,
compare and match relevant image features. In this way, different images of
the same geographical location can be matched. In earth science, geolocating
can be useful to match datasets of geographical areas recorded at different
points in time, by different instruments, with different camera viewpoint,
orientation, resolution and brightness.

For this use case, we process aerial and satellite panchromatic imagery. The
aerial imagery was taken in 2000 and provided by the U.S. Antarctic Resource
Center (USARC). The satellite imagery of the same area was taken in 2017 by
WorldView 2 (WV02) and provided by DigitalGlobe Inc. The geolocating process
involves two main operations: image matching and rectifying of false positive
and false negative matching. For the former we used the scale-invariant
feature transform (SIFT) algorithm~\cite{Lowe2004} and for the latter the
random sampling consensus (RANSAC) algorithm~\cite{bolles1981ransac}.

SIFT is a feature detection algorithm developed for computer vision to detect
commonalities among different images with a stated degree of accuracy and
number of probable false matches. Importantly, SIFT results are invariant to
image resizing and rotation, and partially invariant to changes in brightness
and camera viewpoint. RANSAC is an iterative method to detect outliers in a
provided dataset. It is a ``learning'' algorithm because it fits a model to
multiple random samples of the dataset and returns the model that best fits a
subset of the data. In this context, it is used to evaluate the set of
matched features produced by SIFT and to separate false positive matches.

For our use case, image matching required us to first divide every satellite
image into smaller rectangular tiles of the same size. This process created a
set of tiles between 2,000$^2$~px and 5,000$^2$~px, discarding tiles that
were at the edge of an image. Every tile from one satellite image was then
matched against all the aerial images to find overlapping keypoints, i.e.,
common features such as edges, corners, blobs/regions, and ridges. The
similarity between source and target images was measured as the total number
of matched keypoints, as extracted by SIFT\@.

\section{GPU-SIFT Implementation And Performance Characterization}\label{sec:gpu_imp}

Currently, two main implementations of SIFT are freely available:
CPU-SIFT~\cite{rodriguezs2017fast} and
CUDA-SIFT~\cite{bjorkman2014detecting}. As required by the Geolocation use
case described in~\S\ref{ssec:geoloc-uc}, CPU-SIFT supports raw GeoTIFF
satellite imagery and can process image tiles up to 5,000$^2$~px or more.
Unfortunately, CPU-SIFT is memory inefficient, especially with large tiles,
and cannot use GPUs. CUDA-SIFT supports GPUs but does not support raw GeoTIFF
satellite imagery and can process tiles only up to 1,920$\times$1,080~px.

We addressed these shortcomings by developing
GPU-SIFT~\cite{alsaadi2020scalable}, a novel GPU-based implementation of
SIFT\@. GPU-SIFT offers the following functionalities: (i) reading dual-band
8 and 16 bit GeoTIFF satellite imagery; (ii) reading GeoTIFF images larger
than 1,920$\times$1,080~px; (iii) enabling CUDA to allocate up to 4~GB GPU
memory per image; (iv) implementing CUDA Multi-Process Service (MPS)
technology to run 2 CUDA kernels on single GPU device and (v) implementing
adaptive contrast enhancer.

We characterize and compare the performance of CPU-SIFT, CUDA-SIFT, and
GPU-SIFT based on three metrics: throughput, memory consumption and matching
accuracy. We match single pairs of increasingly large GeoTIFF images (source
and target), measuring how many MB are processed by the CPU or GPU per
second, how much memory the matching of the images required and how accurate
such a matching was. Note that, in this context, throughput refers to the volume of 
data processed per unit of time (MB/s).

We performed all our experiments on the XSEDE Bridges
supercomputer~\cite{nystrom2015bridges}. Bridges offers 32 nodes with 2 Tesla
P100 GPUs and 32~GB of GPU-dedicated memory, and 2 16-cores Intel Broadwell
E5--2683 CPUs with 128~GB of RAM\@. We used
RADICAL-Pilot~\cite{merzky2018using} to manage the execution of our
experiment workloads on Bridges. For all the experiments, we use a single
GeoTIFF image of 845~MB, tiled into four predefined sizes: 90, 180, 360,
720~MB\@. We then match two copies of the same tile for each size, using each
SIFT implementation.

\subsection{Throughput}

We measure the throughput of CPU-SIFT, CUDA-SIFT, and GPU-SIFT in MB/s when
processing the same pair of 90, 180, 360 and 720~MB images. CPU-SIFT and
CUDA-SIFT can analyze one pair of images per CPU/GPU, while GPU-SIFT can use
CUDA MPS to analyze two pairs of images per GPU\@. Consistently, in our
experiments we concurrently execute two CPU-SIFT on two CPUs, two CUDA-SIFT
on two GPUs and one GPU-SIFT on one GPU\@.

\begin{figure}
	\centering
	\begin{subfigure}{0.45\textwidth}
		\includegraphics[width=\linewidth]{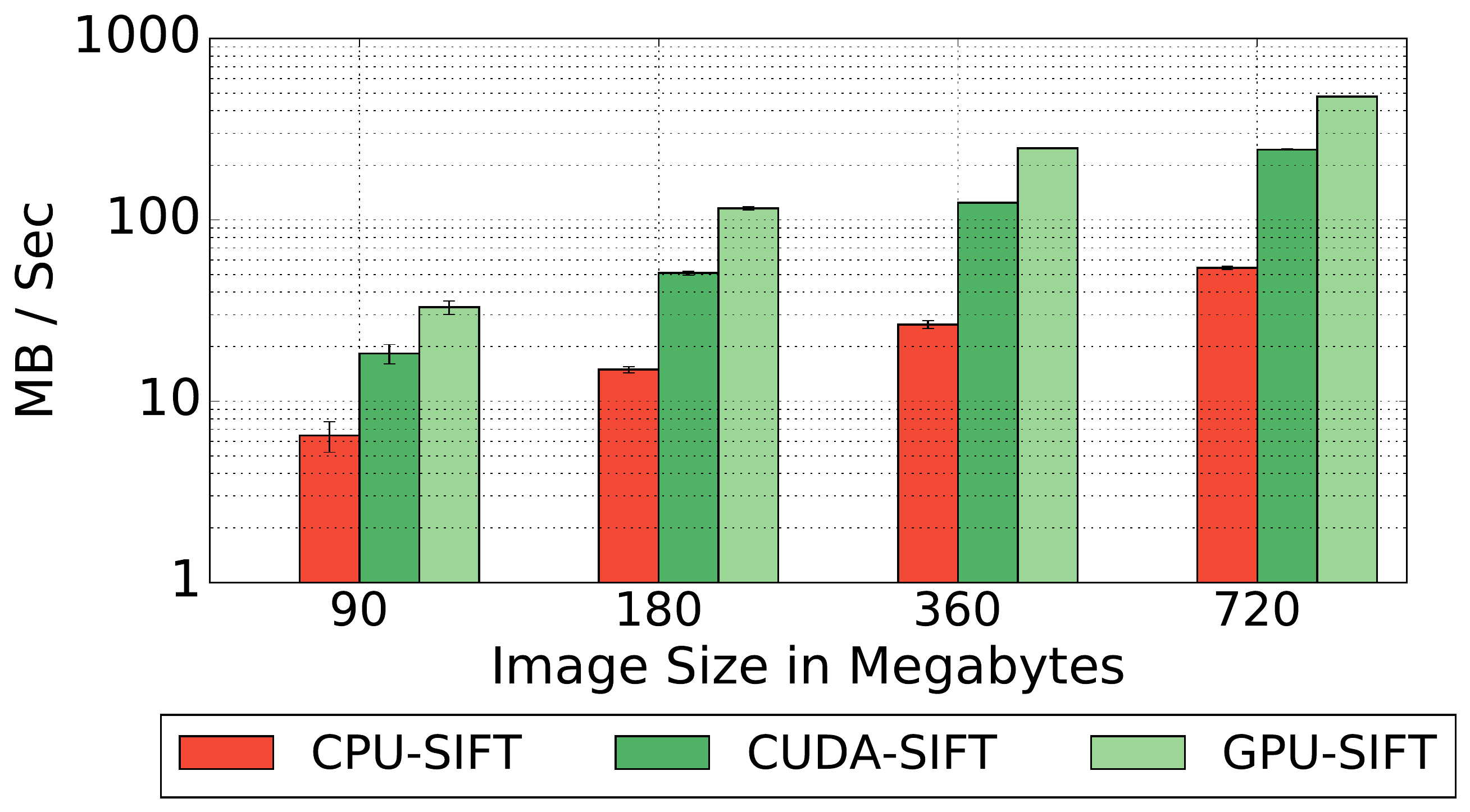}
		\caption{}\label{fig:throughout}
	\end{subfigure}
	~
	\begin{subfigure}{0.45\textwidth}
		\includegraphics[width=\linewidth]{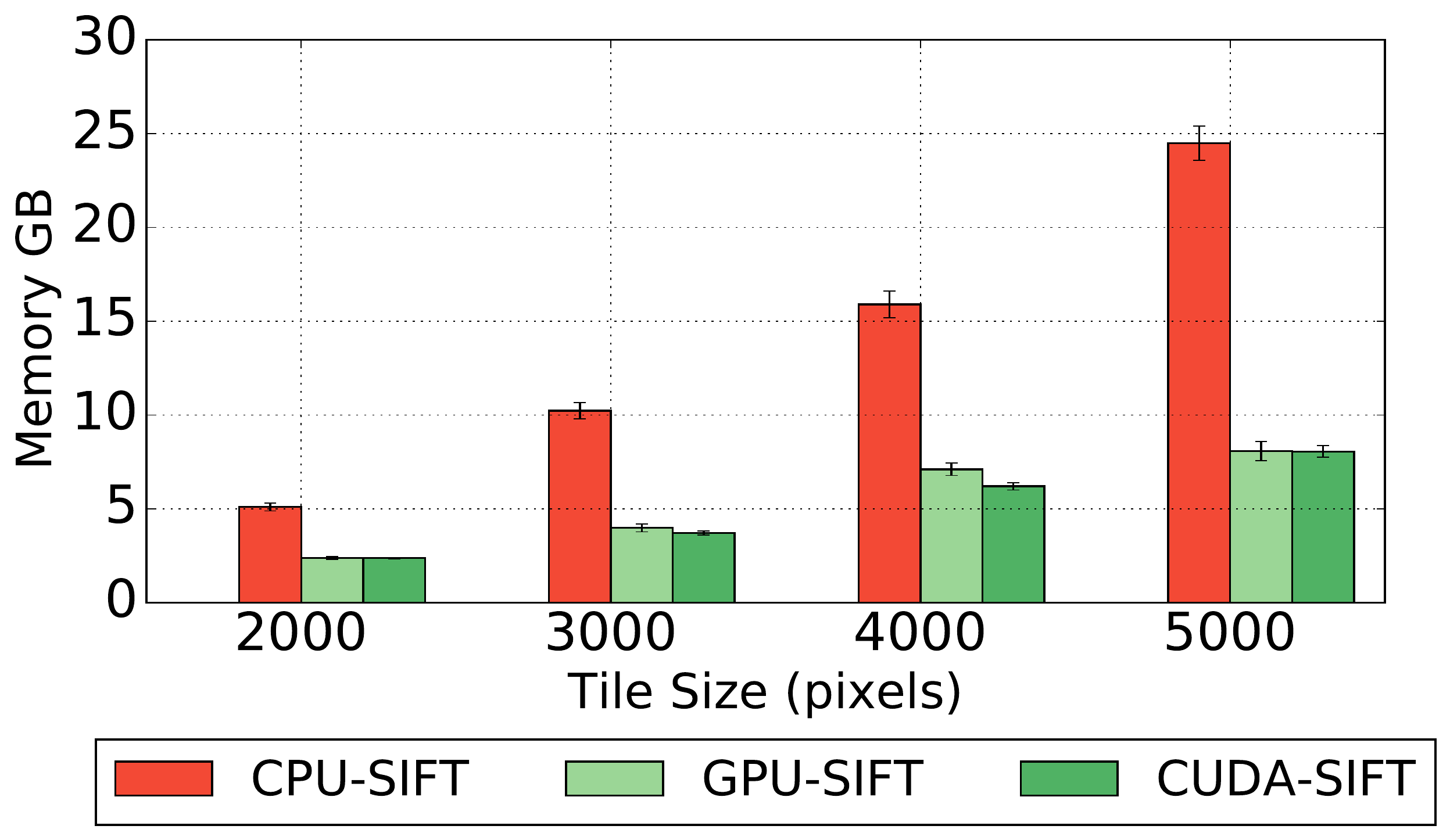}
		\caption{}\label{fig:memory}
	\end{subfigure}
	\caption{(a) The throughput of CPU-SIFT, CUDA-SIFT and GPU-SIFT as a
		function of image size in Megabytes. (b) Memory consumption of
		CPU-SIFT, CUDA-SIFT and GPU-SIFT as a function of tile size.}
\end{figure}

Fig.~\ref{fig:throughout} shows that CPU-SIFT (red) has the lowest throughput
with a value of 54.38~MB/sec. This is due to CPU-SIFT programming model:
CPU-SIFT uses one core per CPU to process and match two images. As a result,
CPU-SIFT cannot leverage the parallelism supported by multi-core architecture
to increase throughput.

CUDA-SIFT (dark green) has a throughput almost six times higher than
CPU-SIFT, with a value of 244.44~MB/sec. In contrast to CPU-SIFT, CUDA-SIFT
uses the CUDA framework to load balance calculations across the cores of the
GPU device. Nonetheless, note that with the given images, CUDA-SIFT uses only
up to 1,658 of the 3,584 CUDA cores available on the Nvidia P100 GPU devices
of our experiments. There are not enough data to process to saturate the GPU
cores.

GPU-SIFT (light green) has the highest throughput among the three SIFT
implementations, with value of 478.25~MB/sec. GPU-SIFT uses 3,400 of the
3,584 available CUDA cores, running 2 pair image comparisons concurrently.
Note that GPU-SIFT throughput is almost double that of the CUDA-SIFT\@. This
shows that the overheads imposed by the CUDA Multi-Process Service (MPS) are
negligible for our implementation.

Fig.~\ref{fig:throughout} shows that GPU throughput is invariant to image
size. Increasing the image size increases the amount of data processed by the
GPU or CPU per second, bounded by the available GPU or CPU bandwidth. Note
that the error bars for the GPU implementations are smaller than those of the
CPU implementation. This is likely due to the efficiency of the CUDA
framework and the absence of competing processes on the GPU subsystem.

\subsection{Memory consumption}

We measure the memory consumption of CPU-SIFT, CUDA-SIFT and GPU-SIFT as the
amount of physical memory a particular program utilizes at runtime.
Fig.~\ref{fig:memory} shows the total memory consumption of CPU-SIFT,
GPU-SIFT and CUDA-SIFT for pairs of 2,000$^2$~px, 3,000$^2$~px, 4,000$^2$~px
and 5,000$^2$~px tiles. The memory usage includes tile reading and the SIFT
detecting, extracting, and matching of features.

For the largest tile of 5,000$^2$~px, CPU-SIFT consumes 24.67~GB of memory,
almost five times higher than CUDA-SIFT and GPU-SIFT which consume 8.05~GB
and 8.08~GB respectively. The minimal difference in memory consumption of
GPU-SIFT compared to CUDA-SIFT account for the acquisition of GeoTIFF images
and the use of CUDA MPS\@. This shows the minimal memory overhead that these
features require and therefore the minimal cost of doubling the throughput of
GPU-SIFT\@.

\subsection{Numbers of Matched points}

We measure the efficiency of each SIFT implementation in terms of the number
of matching points between a pair of images. We picked a pair of images for
which 16,850 matches have been previously identified and validated. We
applied CPU-SIFT, GPU-SIFT and CUDA-SIFT on the pair of images with different
tile sizes to: (1) measure the number of matches that the three
implementations can detect; and (2) validate the assumption that the tile
size can affect the number of matches.

\begin{figure}[ht!]
	\centering
	\begin{subfigure}{0.45\textwidth}
		\includegraphics[width=\linewidth]{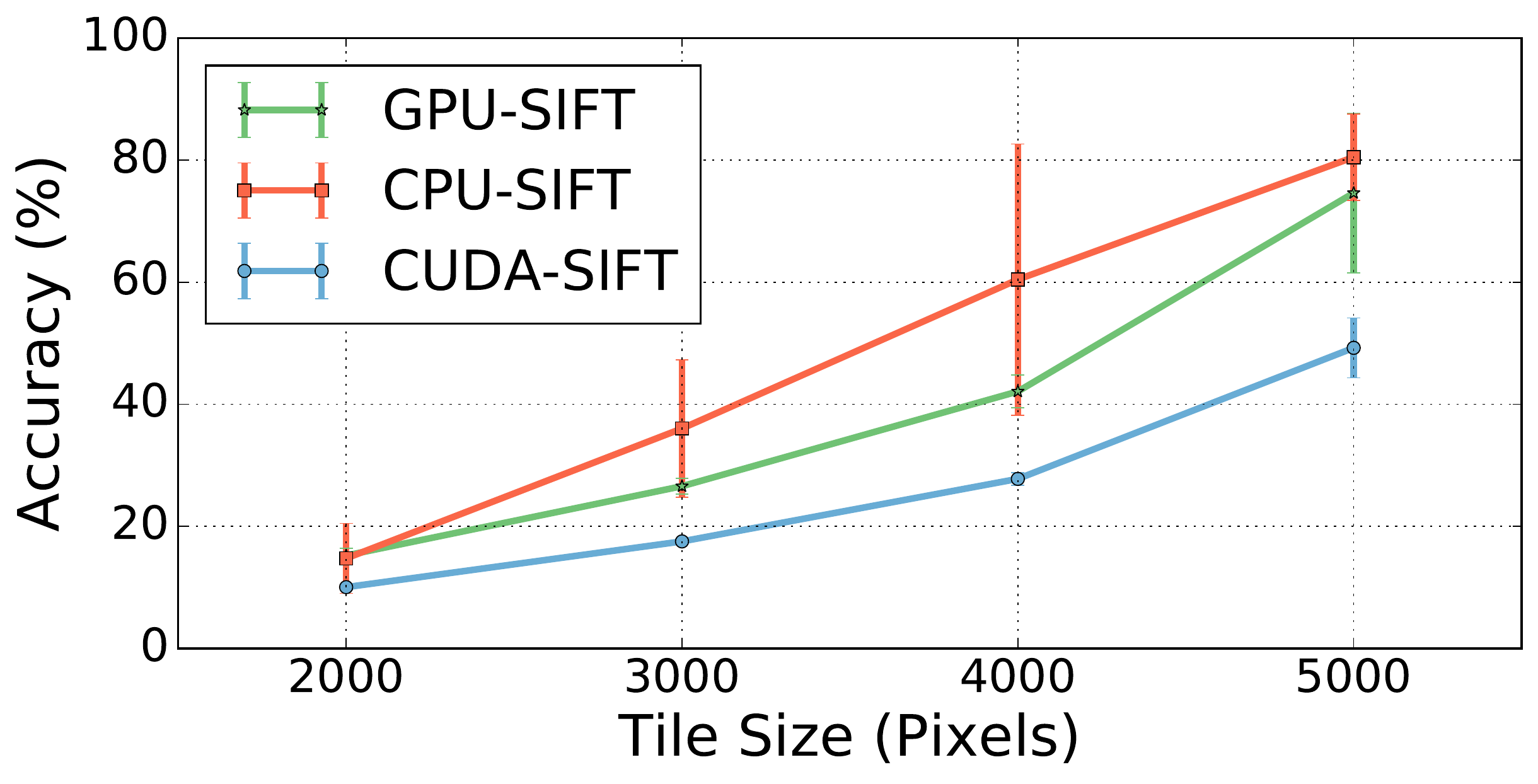}
		\caption{}
		\label{fig:accuracy}
	\end{subfigure}
	~
	\begin{subfigure}{0.45\textwidth}
		\includegraphics[width=\linewidth]{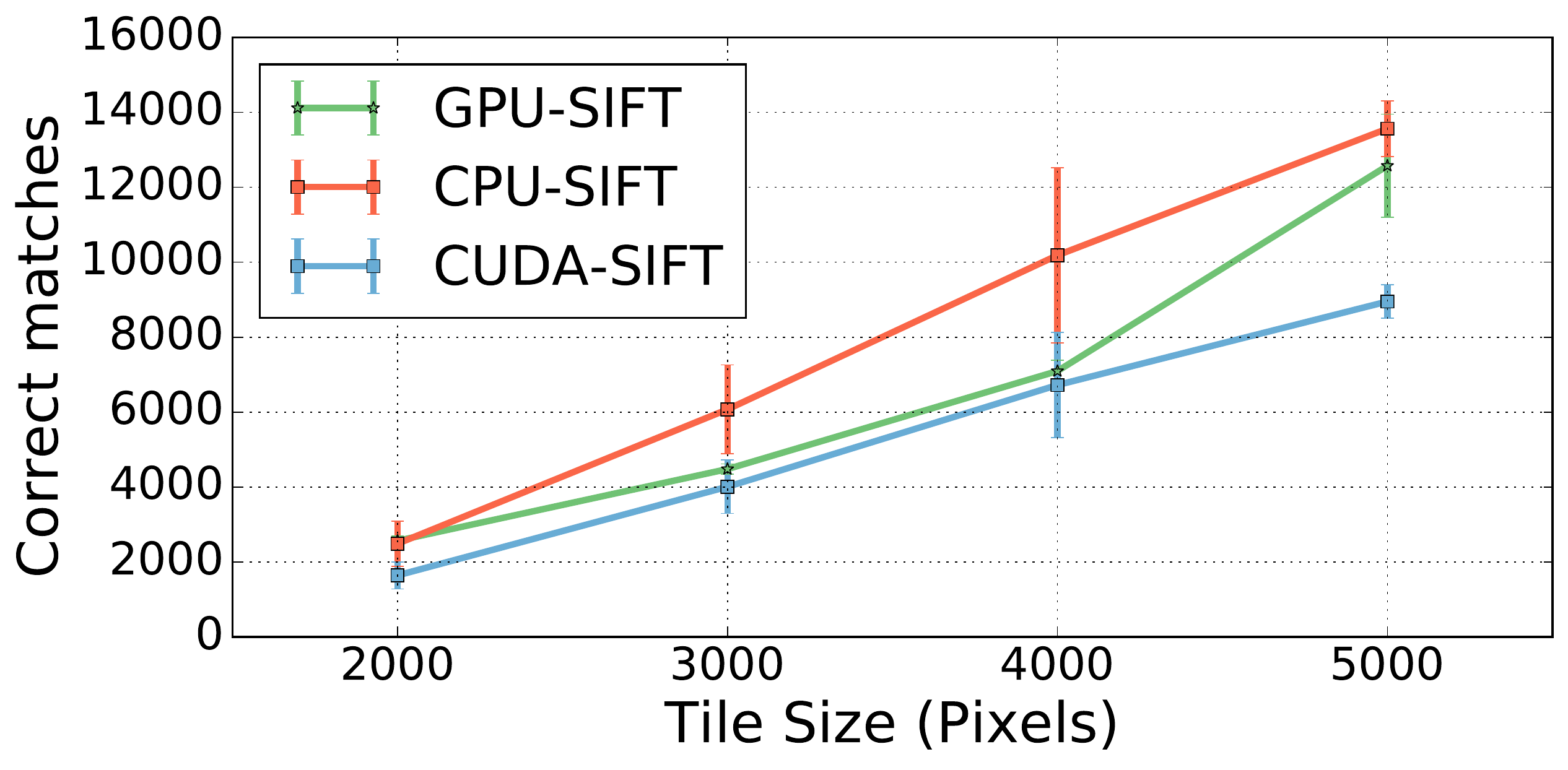}
		\caption{}
		\label{fig:correct_matches}
	\end{subfigure}
	\caption{shows the matching accuracy (a) and the total number of correct
		correspondences (b) found with GPU-SIFT, CUDA-SIFT and CPU-SIFT for
		well-known satellite image as a function of tile
		size.}\label{fig:accuracy_level}
\end{figure}

We repeat the experiments up to 75 times to measure the accuracy of both
implementations, using the same satellite image with tile sizes of
2,000$^2$~px, 3,000$^2$~px, 4,000$^2$~px and 5,000$^2$~px.
Fig.~\ref{fig:accuracy} shows that the number of matched points detected by
CPU-SIFT is more than those detected by GPU-SIFT and CUDA-SIFT\@. For
5,000$^2$~px tiles, CPU-SIFT returns 13,500 matches out of the existing
16,850 with an accuracy of 80.11\%. This is about 5.92\% more than GPU-SIFT,
and 8.97\% more than CUDA-SIFT\@. GPU-SIFT, plotted in green, returns 12,500
matches with an accuracy of 74.18\%, while CUDA-SIFT, plotted in orange,
returns 9,500 matches with an accuracy of 56.81\%.

CPU-SIFT variability shown in Fig.~\ref{fig:accuracy} are larger than those
of CUDA-SIFT and GPU-SIFT and it remains unclear why CPU-SIFT accuracy is
inconsistent. In absolute terms, CPU-SIFT is the implementation that can
reach the highest number of matches but, on average, GPU-SIFT offer a more
reliable performance.

It's worth mentioning that CPU-SIFT and GPU-SIFT apply contrast enhancement
to both source and target images before processing them, while CUDA-SIFT does
not apply any contrast enhancement. Enhancing the level of contrast of an
image can increase the numbers of detected features and as a result, it
directly increases the number of matches between both
images~\cite{Tu2013histogram}.

\section{Workflow Design and Implementation}\label{sec:w-design}

Computationally, the use cases described in~\S\ref{sec:ucase} present three
main challenges: heterogeneity, scale and reusability. The images of the
use cases' datasets have a wide distribution in size. Each image
requires a series of tasks to get the aggregated result. These tasks are
memory and computational intensive, requiring CPU and GPU implementations.
Whenever the image dataset is updated, it needs to be reprocessed.

We address these challenges by codifying image analyses into workflows. We
then execute these workflows on HPC resources, leveraging the concurrency,
storage systems and compute speed they offer to reduce time to completion.
Typically, the workflows of our use cases consist of a sequence (i.e.,
pipeline) of tasks, each performing part of the end-to-end analysis on one or
more images. We compare two common designs for the execution of these
workflows: one in which each image is processed independently by a dedicated
pipeline, and the other in which a single pipeline processes multiple images.

Note that both designs separate the functionalities required to process each
image from the functionalities used to coordinate the processing of multiple
images. This is consistent with moving away from vertical, end-to-end
single-point solutions, favoring designs and implementations that satisfy
multiple use cases, possibly across diverse domains. Accordingly, the designs we implement and characterize, employ two tasks (i.e., standalone executable programs) to provide the functionalities required by the use cases.

The designs are functionally equivalent, in that they both enable the
analysis of the given image datasets. Nonetheless, each design leads to
different concurrency, resource utilization and overheads, depending on
compute-data affinity, scheduling algorithms, and coordination between CPU
and GPU computations. We analyze the performance of these designs using the
common metrics of total execution time, resource utilization, and middleware
overheads.

Consistent with HPC resources currently available and our use
cases, we make three assumptions:
\begin{inparaenum}[(1)]
    \item each compute node has $c$ CPUs;
    \item each compute node has $g$ GPUs where $g \le c$; and
    \item each compute node has enough memory to enable concurrent execution of
    a certain number of tasks.
\end{inparaenum}
As a result, at any given point in time there are $C = n\times c$ CPUs and $G
= n\times g$ GPUs available, where $n$ is the number of compute nodes.

\begin{figure*}[ht!]
    \centering
    \begin{subfigure}[b]{0.32\textwidth}
        \includegraphics[width=\linewidth]{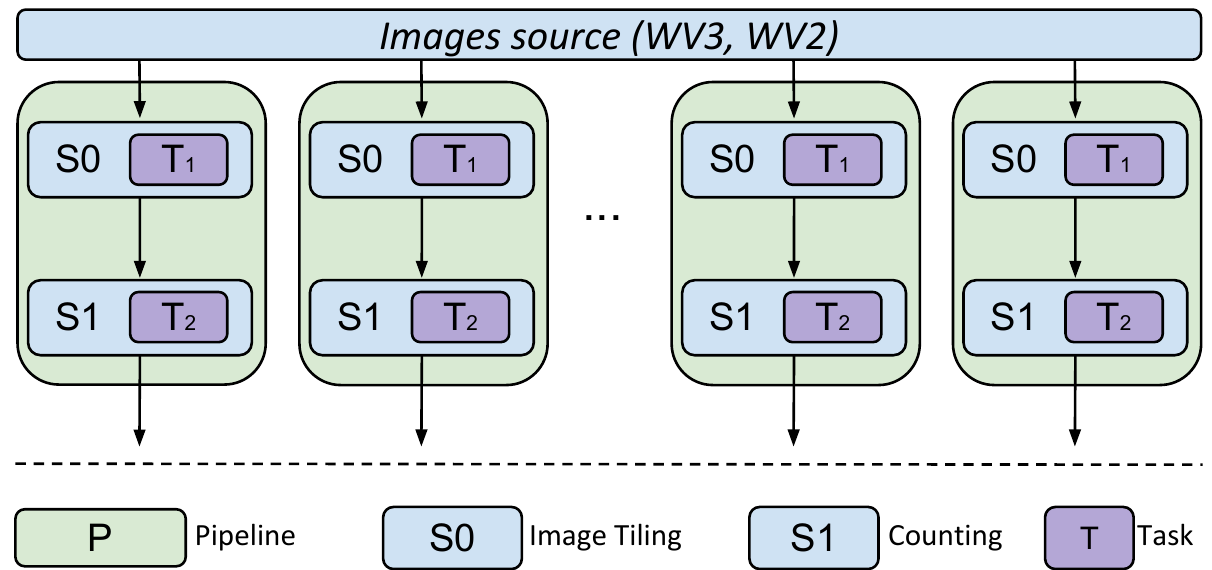}
        \caption{}
		\label{fig:seals_design1}
    \end{subfigure}%
    ~ 
    \begin{subfigure}[b]{0.32\textwidth}
        \includegraphics[width=\linewidth]{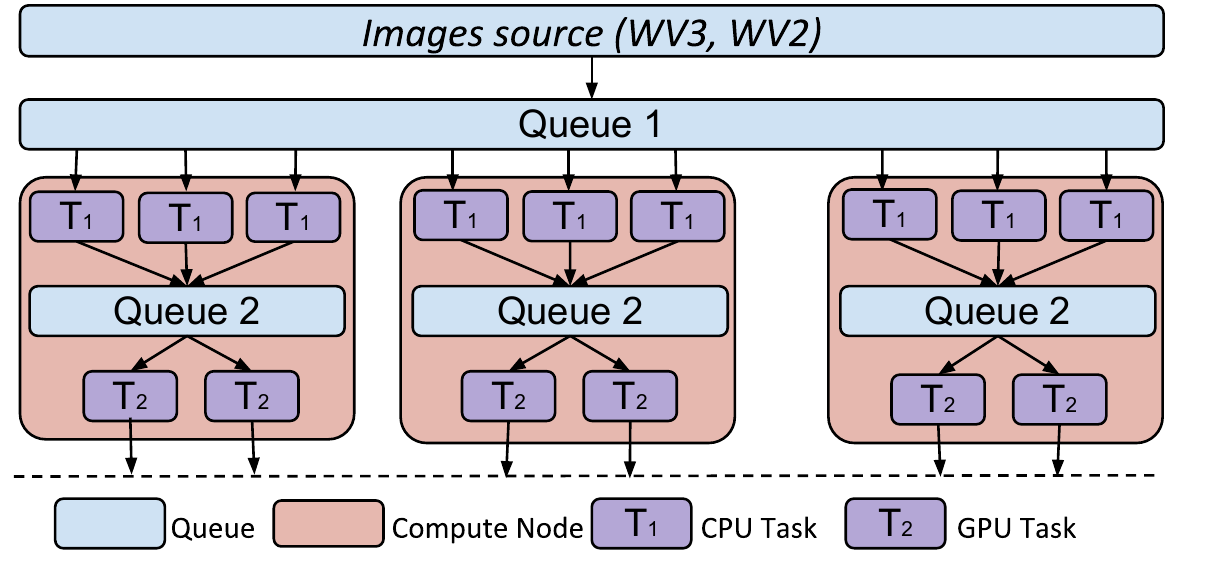}
        \caption{}\label{fig:seals_design2}
    \end{subfigure}%
    ~ 
    \begin{subfigure}[b]{0.32\textwidth}
        \includegraphics[width=\linewidth]{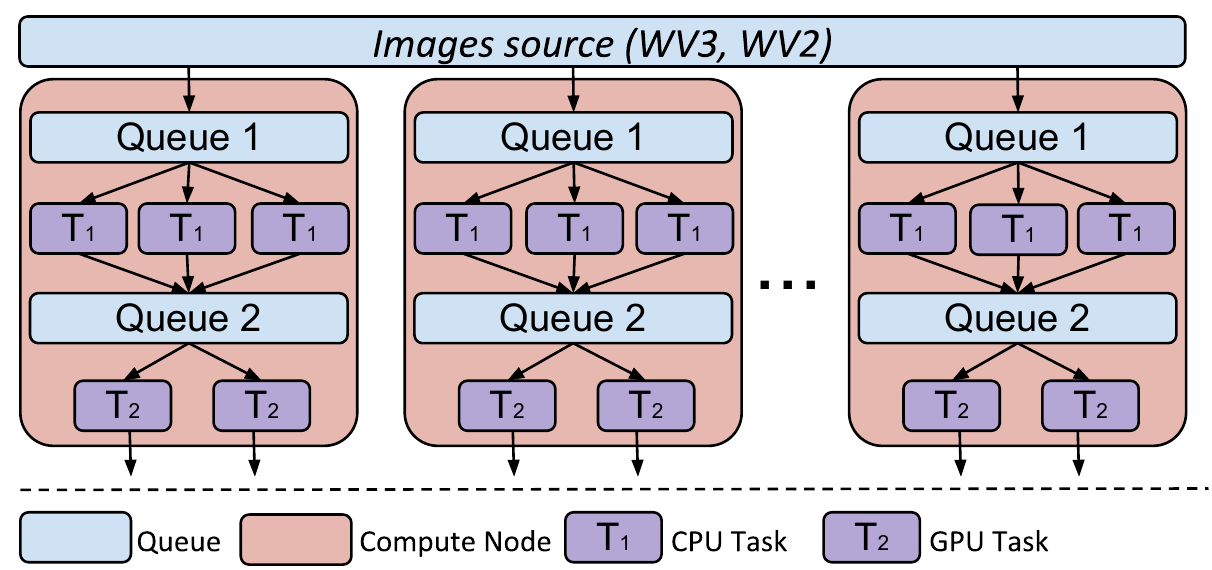}
		\caption{}\label{fig:seals_design3}
    \end{subfigure}
    \caption{Design approaches to implement the workflow required for the
    use cases of~\S\ref{sec:ucase}.~\ref{fig:seals_design1}--\textbf{Design
    1}: Pipeline, stage and task based
    design.~\ref{fig:seals_design2}--\textbf{Design 2}: Queue based design
    with a single queue for all the tasks of the
    pipeline.~\ref{fig:seals_design3}--\textbf{Design 2.A}: Queue based
    design with multiple queues for the tasks of the
    pipeline.}\label{fig:designs}
\end{figure*}

\subsection{Design 1: One Image per
Pipeline}\label{ssec:approach1}\label{des1}

We specify the workflow for either counting the number of seals in a set of
images, or geolocating pairs of images as a set of pipelines. Each pipeline
is composed of two stages, each with multiple instances of one type of task.
In the Seals use case (UC1), we specify the workflow for counting the number
of seals in a set of images. In the Geolocation use case (UC2), we specify
the workflow to match a pair of images in two sets of aerial and satellite
images.

In UC1, the task of the first stage gets an image as input and generates
tiles of that image based on the tile size as output. In UC2, the task of the
first stage gets a pair of images as input and produces a set of matches
between these two images as output. The task of the second stage gets the
output of the first stage---generated tiles or matches---and outputs the the
number of seals per image for UC1 and a set of matches with reduced false
positives for UC2.

Formally, we define two types of tasks in UC1 and UC2:

\begin{itemize}
    \item $T^{UC1}_{1} = <I, f_{I}, t>$, where $I$ is an image or a pair of
    images, $f_{I}$ is a stage 1 type function and $t$ is a set of tiles or
    matches that correspond to $I$.
    \item $T^{UC1}_{2} = <t, f_{A}, S>$, where $f_{A}$ is a stage 2 type
    function and $S$ is either the number of seals or final matches.
    \item $T^{UC2}_{1} = <I_{p}, k_{I}, m>$, where $I_{p}$ is a pair of
    images, $k_{I}$ is a matching function and $m$ is a set of matches as a
    file that correspond to $I_{p}$.
    \item $T^{UC2}_{2} = <m, f_{R}, t>$, where $f_{R}$ is a function that
    eliminates the undesired matches and $t$ is the output file of filtered
    matches.
\end{itemize}

Tiling in $T^{UC1}_{1}$ is implemented with OpenCV~\cite{bradski2000opencv}
and Rasterio~\cite{gillies2013rasterio} in Python. Rasterio allows us to open
and convert a GeoTIFF WV3 image to an array. The array is then partitioned to
sub-arrays based on a user-specified scaling factor. Each sub-array is
converted to an compressed image via OpenCV routines and saved to the
filesystem.

Seal counting in $T^{UC1}_{2}$ is performed via a Convolutional Neural
Network (CNN) implemented with PyTorch~\cite{paszke2017automatic}. The CNN
counts the number of seals for each tile of an input image. When all tiles
are processed, the coordinates of the tiles are converted to the geographical
coordinates of the image and saved in a file, along with the number of
counted seals. Note that the number of seals in an tile does not affect the
execution of the network, i.e., the same number of operations will be
executed.

Matching in $T^{UC2}_{1}$ is implemented by GPU-SIFT as described
in~\S\ref{sec:gpu_imp} while filtering in $T^{UC2}_{2}$ is implemented by
RANSAC as described in~\ref{ssec:geoloc-uc}.

All task implementations for UC1 and UC2 are invariant across the alternative
designs we consider. This is consistent with the designs being task-based,
i.e., each task exclusively encapsulates the capabilities required to perform
a specific operation over an image, pair of images or tile. Thus, tasks are
independent from the capabilities required to coordinate their execution,
whether each task processes a single image or pair of images, or a sequence
of images or pairs of images.

We implemented Design~1 via \entk, a workflow engine which exposes an API
based on pipelines, stages, and tasks~\cite{balasubramanian2018harnessing}.
The user can define a set of pipelines, where each pipeline has a sequence of
stages, and each stage has a set of tasks. Stages are executed respecting
their order in the pipeline while the tasks in each stage can execute
concurrently, depending on resource availability.

For our use cases, \entk{} has three main advantages compared to other
workflow engines:
\begin{inparaenum}[(1)]
    \item it exposes pipelines and tasks as first-order abstractions
    implemented in Python;
    \item it is specifically designed for concurrent management of up to
    $10^5$ pipelines; and
    \item it supports RADICAL-Pilot, a pilot-based runtime system designed to
    execute heterogeneous bag of tasks on HPC
    machines~\cite{merzky2018using}.
\end{inparaenum} 
Together, these features address the challenges of heterogeneity, scale and
reusability: users can encode multiple pipelines, each with different types
of tasks, executing them at scale on HPC machines without explicitly coding
parallelism and resource management.

When implemented in \entk, the workflows of our use cases map to a set of
pipelines, each with two stages $St_{1}$, $St_{2}$. Each stage has a task of
type $T^{UC1-2}_{1}$ and $T^{UC1-2}_{2}$ respectively. Each pipeline is
defined as $P = (St_{1},St_{2})$. For UC1, the workflow consists of $N$
pipelines and for UC2 the workflow consists of $N\times (N-1)$, where $N$ is the
number of images. 

Figure~\ref{fig:seals_design1} shows the abstract workflow for both use
cases. For each pipeline, \entk{} submits the task of stage $St_{1}$ to the
runtime system (RTS). As soon as this task finishes, the task of stage
$St_{2}$ is submitted for execution. This design allows concurrent execution
of pipelines and, as a result, concurrent analysis single images or pair of
images, one by each pipeline. Since pipelines execute independently and
concurrently, there are instances where $St_{1}$ of a pipeline executes at
the same time as $St_{2}$ of another pipeline.

Design~1 has the potential to increase utilization of available resources as
each compute node of the target HPC machine has multiple CPUs and GPUs.
Importantly, computing concurrency comes with the price of multiple reads and
writes to the filesystem on which the dataset is stored. This can cause I/O
bottlenecks, especially if each task of each pipeline reads from and writes to
the same filesystem, possibly over a network connection.

For UC1, we used a tagged scheduler for \entk's RTS to avoid I/O bottlenecks.
This scheduler schedules $T_{1}$ of each pipeline on the first available
compute node, and guarantees that the corresponding $T_{2}$ is scheduled on
the same compute node. As a result, compute-data affinity is guaranteed among
co-located $T_{1}$ and $T_{2}$. This design reduces I/O bottlenecks but it may
also reduce concurrency when the performance of the compute nodes and/or the
tasks is heterogeneous: $T_{2}$ may have to wait to execute on a specific
compute node while another node is free.

\subsection{Design 2: Multiple images per pipeline}\label{ssec:approach2}

Design~2 implements a queue-based design. We introduce two tasks ($T_{1}$-
$T_{2}$) for both UC1 and UC2 as defined in~\S\ref{ssec:approach1}. In
contrast to Design 1, these tasks are started and then executed for as long
as resources are available, processing input images at the rate taken to
process each image or pair of images. For both use cases, the number of
concurrent $T_{1}$ and $T_{2}$ depends on available resources, including
CPUs, GPUs, and RAM\@.

For the implementation of Design~2, we do not need \entk{}, as we submit a
bag of $T_{1}$ and $T_{2}$ tasks via the RADICAL\nobreakdash-Pilot RTS, and
manage the data movement between tasks via queues. As shown in
Fig.~\ref{fig:seals_design2}, Design 2 uses one queue (Queue 1) for the
dataset, and another queue (Queue 2) for each compute node. For each compute
node, each $T_{1}$ pulls an image or pair of images from Queue 1, generates
tiles or matches and then queues the results to Queue 2. The first available
$T_{2}$ on that compute node, pulls those tiles or matches from Queue 2, and
counts the seals or filters false positive matches.

To communicate data and control signals between queues and tasks, we defined
a communication protocol with three entities: Sender, Receiver, and Queue.
Sender connects to Queue and pushes data. When done, Sender informs Queue and
disconnects. Receiver connects to Queue and pulls data. If there are no data
in Queue but Sender is connected, Receiver pulls a ``wait'' message, waits,
and pulls again after a second. When there are no data in Queue or Sender is
not connected to Queue, Receiver pulls an ``empty'' message, upon which it
disconnects and terminates. This ensures that tasks are executing, even if
starving, and that all tasks are gracefully terminating when all images are
processed.

Note that Design~2 load balances $T_{1}$ tasks across compute nodes but
balances $T_{2}$ tasks only within each node. For example, suppose that
$T_{1}$ on compute node $A$ runs two times faster than $T_{1}$ on compute
node $B$. Since both tasks are pulling images from the same queue, $T_{1}$ of
$A$ will process twice as many images as $T_{1}$ of $B$. Both $T_{1}$ of $A$
and $B$ will execute for around the same amount of time until Queue 1 is
empty, but Queue 2 of $A$ will be twice as large as Queue 2 of $B$. $T_{2}$
tasks executing on $B$ will process half as many images as $T_{2}$ tasks on
$A$, possibly running for a shorter period of time, depending on the time
taken to process each image.

In principle, Design~2 can be modified to load balance also across Queue 2
but in practice, as discussed in~\S\ref{ssec:approach1}, this would produce
I/O bottlenecks: Load balancing across $T_{2}$ tasks would require for all
tiles produced by $T_{1}$ tasks to be written to and read from a filesystem
shared across multiple compute nodes. Keeping Queue 2 local to each compute
node enables using the filesystem local to each compute node.

\subsubsection{Design 2.A\@: Uniform image dataset per
pipeline}\label{sssec:approach2a}

The lack of load balancing of $T_{2}$ tasks in Design 2 can be mitigated by
introducing a queue in each node from where $T_{1}$ tasks pull data. This
allows early binding of images to compute nodes, i.e., deciding the
distribution of input images per node before executing $T_{1}$ and $T_{2}$.
As a result, the execution can be load balanced among all available nodes,
depending on the correlation between image properties and image execution
time.

Figure~\ref{fig:seals_design3} shows variation 2.A of Design~2. The early
binding of images to compute nodes introduces an overhead compared to using
late binding via a single queue as in Design 2. Nonetheless, depending on the
strength of the correlation between image properties and execution time,
design 2.A offers the opportunity to improve resource utilization. While in
Design 2 some node may end up waiting for another node to process a much
larger Queue 2, in design 2.A this is avoided by guaranteeing that each
compute node has an analogous payload to process.

\section{Experiments and Discussion}\label{sec:experiments}

We executed three experiments using the GPU compute nodes of the XSEDE
Bridges supercomputer. These nodes offer 32 cores, 128 GB of RAM and two P100
Tesla GPUs. We stored the dataset of the experiments and the output files on
Bridges' Pylon5 Lustre filesystem. Specifically, for the Seals use case (UC1
onwards), we stored the tiles produced by the tiling tasks on the local
filesystem of the compute nodes. This way, we avoided a potential performance
bottleneck from millions of reads and writes of $\approx$700~KB on
Pylon5. The Geolocation use case (UC2 onwards) did not require the use of the
node local filesystem since it writes a single file of few MBs per task. We
submitted jobs requesting 4 compute nodes to keep the average queue time
within a couple of days. Requesting more nodes produced queue times in excess
of a week.

The dataset of UC1 consists of 3,097 images, ranging from 50 to 2,770~MB, for
a total of 4~TB of data. The dataset of UC2 consists of of 1,575 aerial and
satellite images, ranging from 1.5 to 5.5~MB for a total of 4.35~GB\@. We
generated 11,552 image pairs to cross-match all aerial images to all
satellite images. The image size of both datasets follows a normal
distribution. The UC1 dataset has a mean value of 1,304.85~MB and standard
deviation of 512.68~MB\@. The dataset of UC2 has a mean value of 6.13~MB and
standard deviation of 1.79~MB\@.

For Design~1, 2 and 2.A described in~\S\ref{sec:w-design}, Experiment~1
models the execution time of the two tasks of our use cases as a function of
the image size---the only property of the images for which we found a
correlation with execution time; Experiment~2 measures resource utilization
for each design; and Experiment~3 characterizes the overheads of the
middleware implementing each design. These experiments enable performance
comparison across designs, allowing us to draw conclusions about the
performance of heterogeneous task-based execution of data-driven workflows on
HPC resources.

As already done in~\S\ref{des1}, we use $T^{UC1}_{1}$ and $T^{UC1}_{2}$ to
indicate the first and second type of task for the Seals use case; and
$T^{UC2}_{1}$ and $T^{UC2}_{2}$ to indicate first and second type of task for
the Geolocation use case.

\subsection{Experiment~1, Design~1:}\label{ssec:des1analysis}

\begin{figure*}[ht!]
    \centering
    \begin{subfigure}[b]{0.49\textwidth}
        \includegraphics[width=\linewidth]{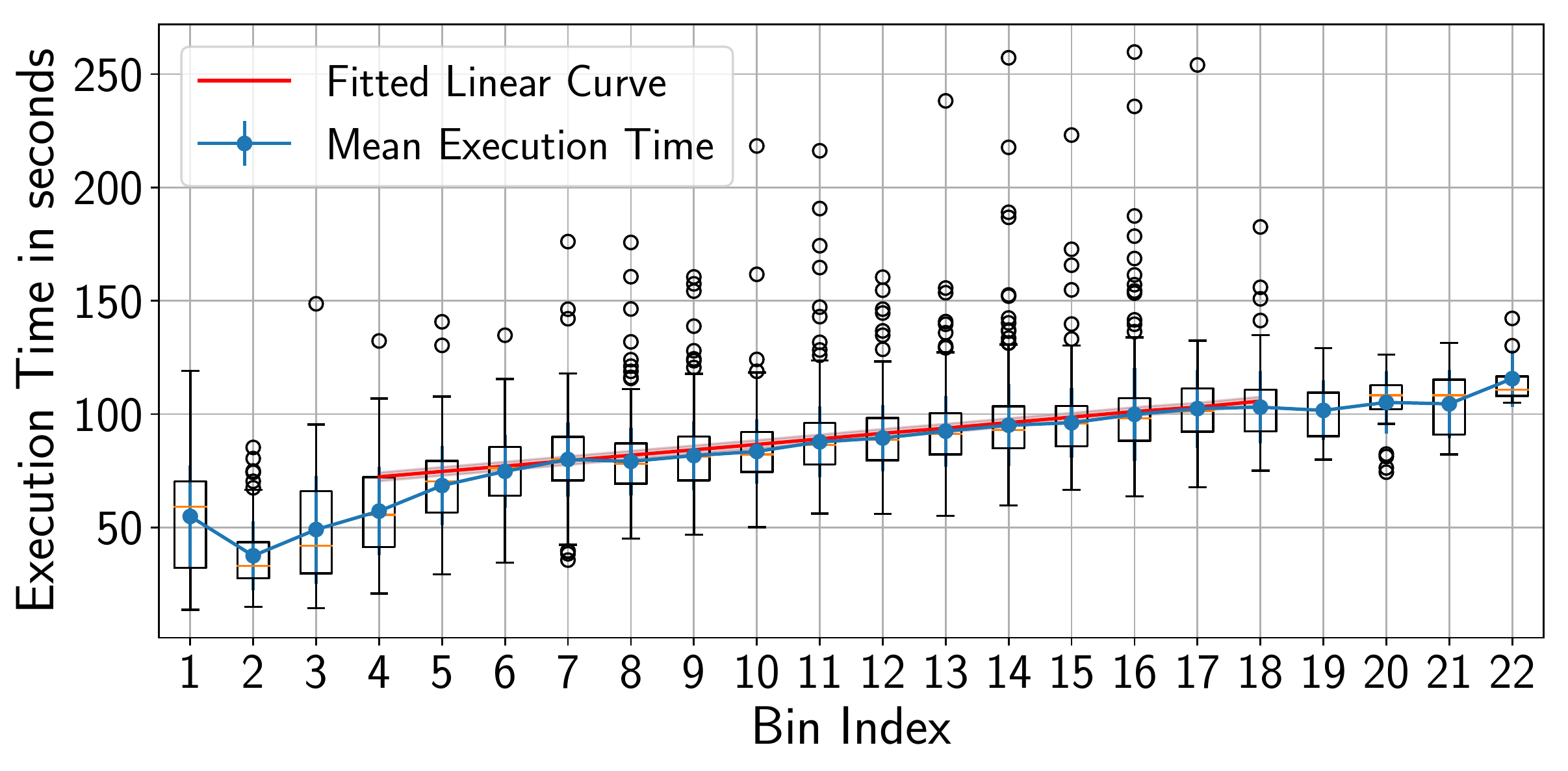}
        \caption{}\label{fig:stage_0_execution}
    \end{subfigure}%
    ~ 
    \begin{subfigure}[b]{0.49\textwidth}
        \includegraphics[width=\linewidth]{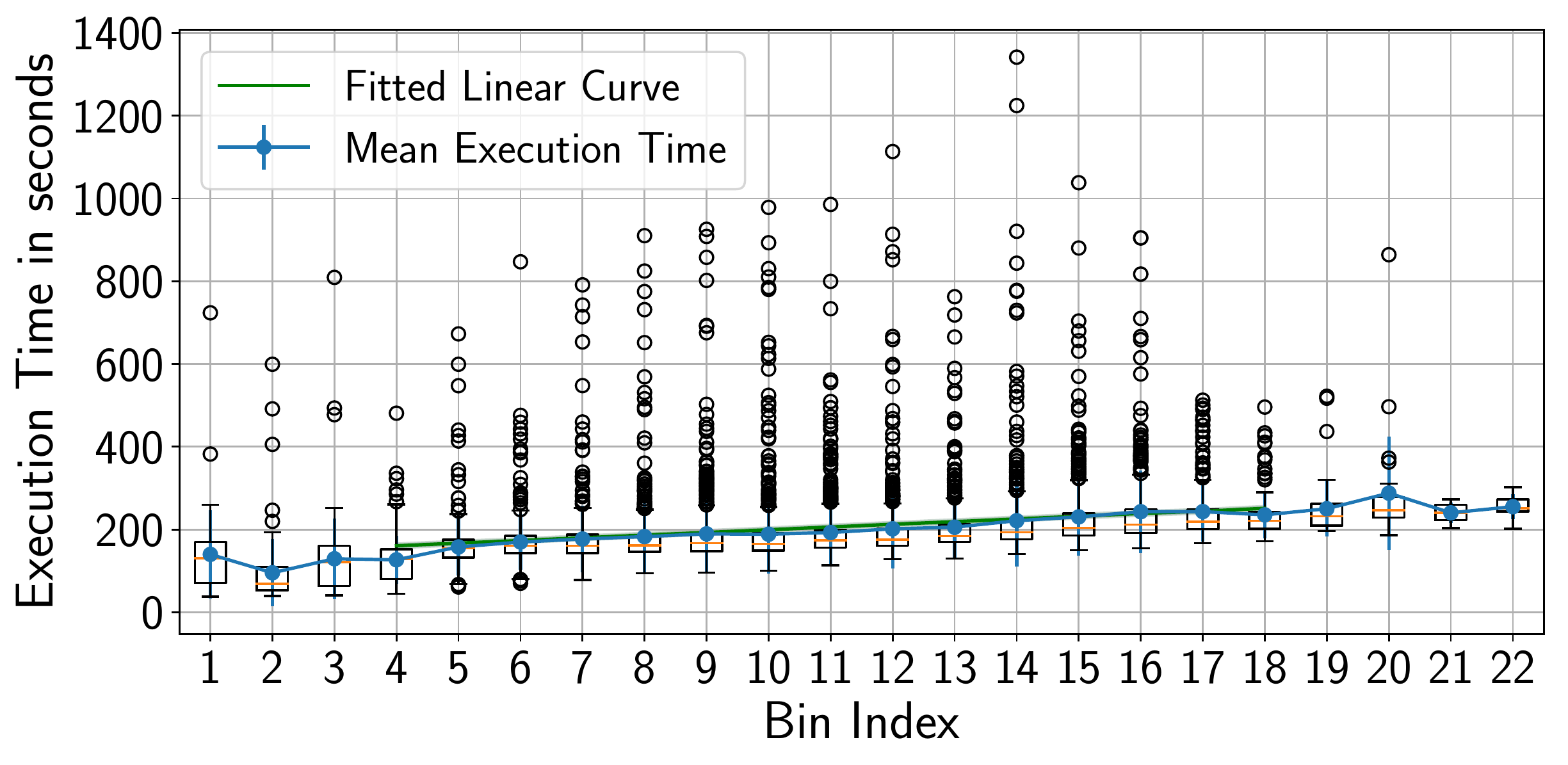}
        \caption{}\label{fig:stage_1_execution}
    \end{subfigure}
    \caption{Experiment~1, Design~1, UC1: Box-plots of (a) 
    $T^{UC1}_{1}$ and (b) $T^{UC1}_{2}$ execution times, means and standard
    deviations ($STDs$) for $125$~MB image size bins. Red line shows fitted
    linear function for $T^{UC1}_{1}$, green line for $T^{UC1}_{2}$. Red
    shadow shows confidence interval for $T^{UC1}_{1}$, green shadow for
    $T^{UC1}_{2}$.}\label{fig:execution}
\end{figure*}

Fig.~\ref{fig:stage_0_execution} shows the execution time of the tiling task
$T^{UC1}_{1}$ as a function of the image size. We partition the set of images
based on image size, obtaining 22 bins with binsize 125~MB each starting from
50~MB up to 2,800~MB\@. The average time to tile an image in each bin tends
to increase with the image size. The box-plots show some positive skew of the
data with a number of data points falling outside the assumed normal
distribution. Thus, there is a weak correlation between task execution time
and image size with a large spread across all the image sizes.

There are also large standard deviations ($STD$---blue line) in most of the
bins. We explored the causes of the observed values by measuring how it
varies in relation to the number of $T^{UC1}_{1}$ concurrently executing on
the same node. The $STD$ observed was consistent across degrees of task
concurrency allowing us to conclude that it depends on fluctuation in the
node performance~\cite{paraskevakos2019workflow}.

\begin{figure*}[ht!]
    \centering
    \begin{subfigure}[b]{0.49\textwidth}
        \includegraphics[width=\linewidth]{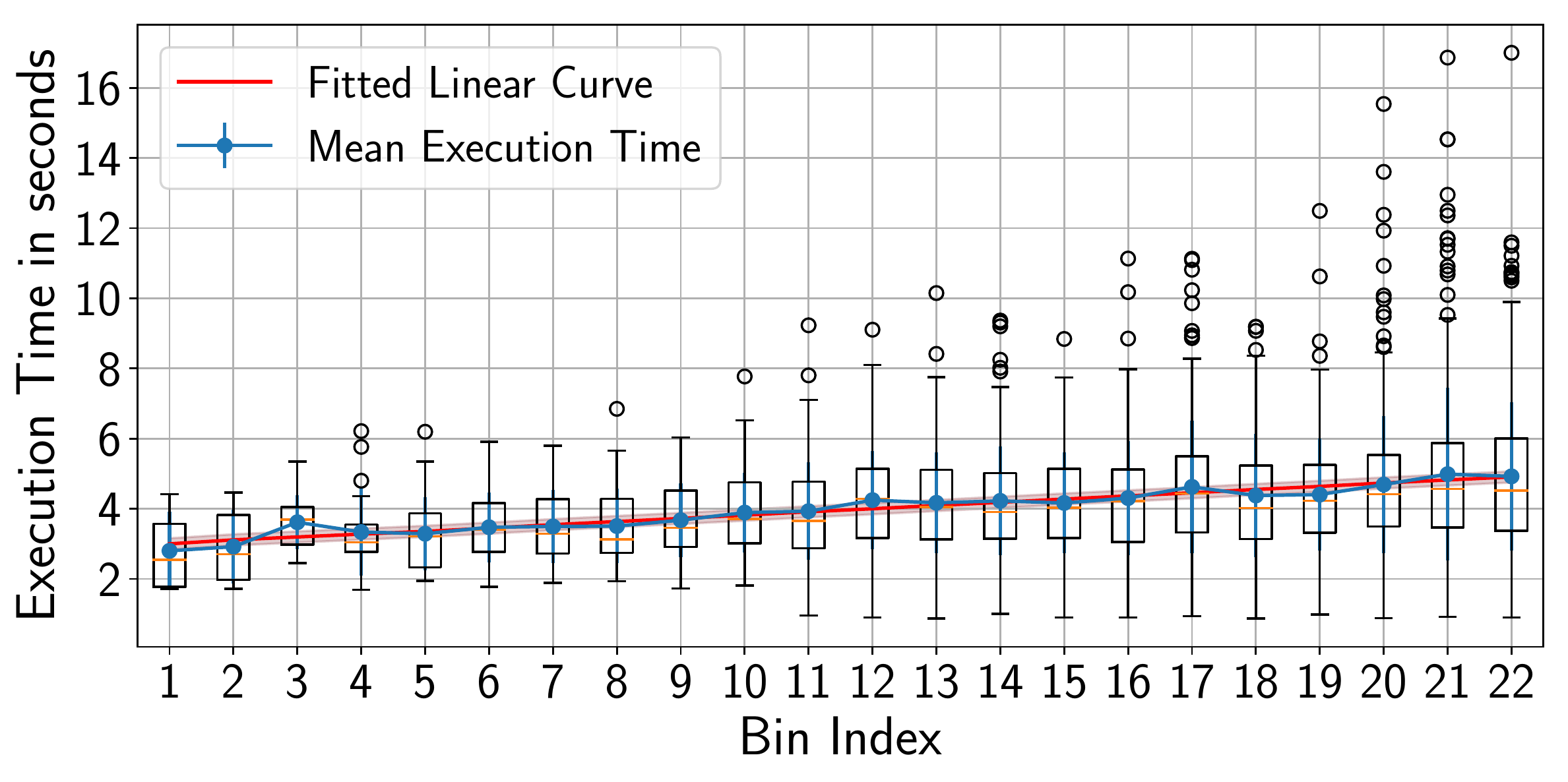}
        \caption{}\label{fig:geo_stage_0_execution}
    \end{subfigure}%
    ~ 
    \begin{subfigure}[b]{0.49\textwidth}
        \includegraphics[width=\linewidth]{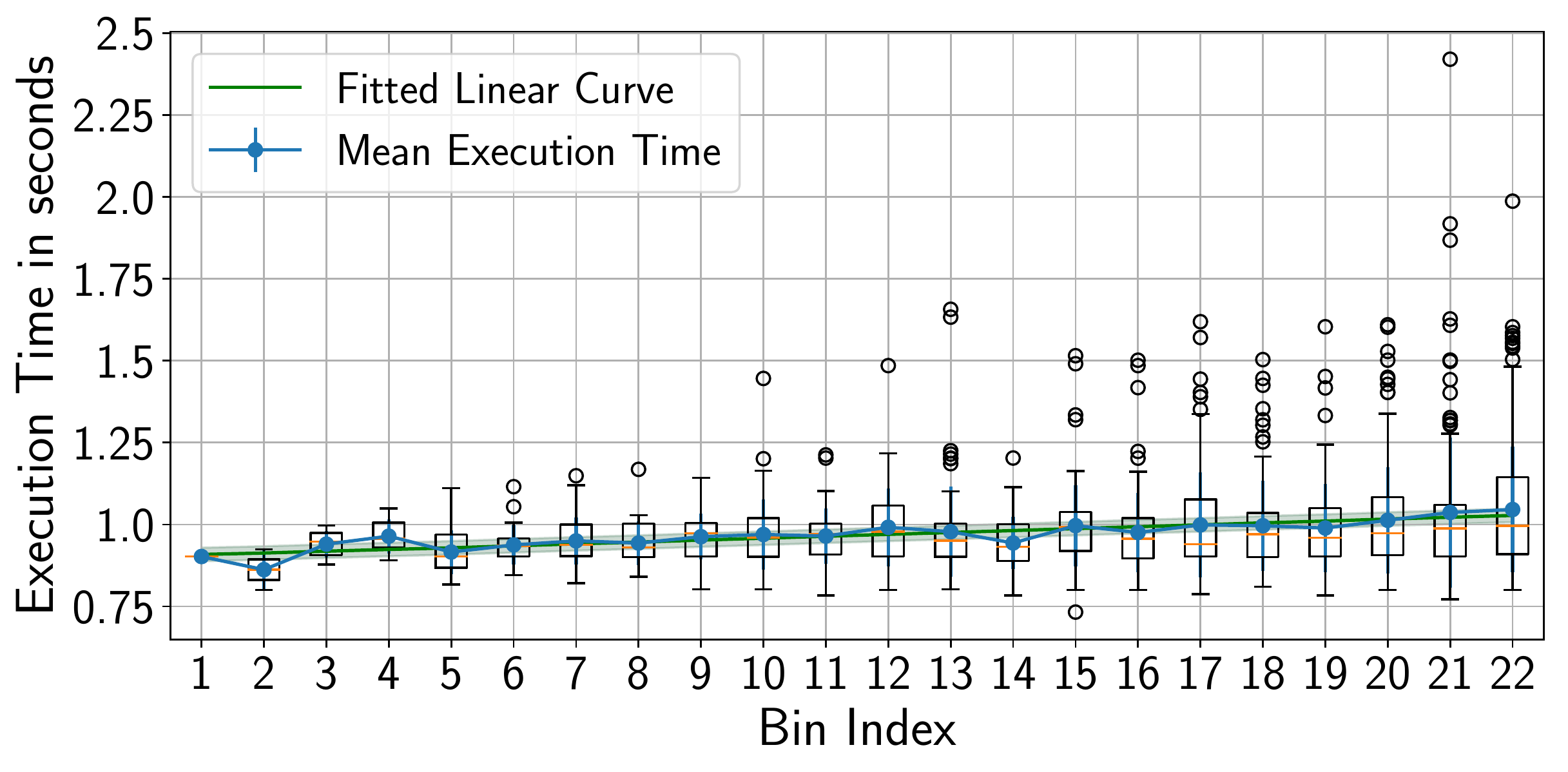}
        \caption{}\label{fig:geo_stage_1_execution}
    \end{subfigure}
    \caption{Experiment~1, Design~1, UC2: Box-plots of (a) $T^{UC2}_{1}$ and
        (b) $T^{UC2}_{2}$ execution times, means and standard deviations
        ($STDs$) for $187$~MB image size bins. Red line shows fitted linear
        function for $T^{UC2}_{1}$, green line for $T^{UC2}_{2}$. Red shadow
        shows confidence interval for $T^{UC2}_{1}$, green shadow for
        $T^{UC2}_{2}$.}\label{fig:geo_execution}
\end{figure*}

Fig.~\ref{fig:geo_stage_0_execution} shows the execution time of the image
matching task $T^{UC2}_{1}$ as a function of the size of an image pair. In
this use case, we partitioned the data based on the total size of the image
pair as each task processes two images at the same time.
Fig.~\ref{fig:geo_stage_0_execution} shows 22 bins of 187~KB, each in a range
of $[1.0,5.5]$~MB\@.

Fig.~\ref{fig:stage_0_execution} indicates that the execution time is a
linear function of the image size between bins 4 and 18. Bins 1--3 and 19--23
are not representative as the head and tail of the image sizes distribution
contain less than 5\% of the image dataset. Similarly, in
Fig.~\ref{fig:geo_stage_0_execution} bins 5--19 show linear behavior and bins
1--4 and 20--23 are omitted from the analysis as they contain less than 4\%
of the data set. Accordingly, for both UC1 and UC2, we model the execution
time as:
\begin{equation}
	T(x) = \alpha \times x+\beta
    \label{eq:des1_til}
\end{equation} where $x$ is the image size. We found the parameter values 
of Eq.~\ref{eq:des1_til} by using a non-linear least squares algorithm to fit
experimental data (see Table~\ref{tab:fit_par_val}).

\begin{table*}[ht]
    \centering
    \scriptsize
    \begin{tabular}{@{}cclrcl@{}}
        \toprule
        \textbf{Design}                                &
        \textbf{Fitted Data}                           &
        \textbf{$\alpha$ value}                        &
        \textbf{$\beta$ value}                         &
        \textbf{$R^2$ value}                           &
        \textbf{Figure}                                \\
        \midrule
        1                                              & 
        $T^{UC1}_{1}$                                  & 
        $1.92\times 10^{-2}$                           & 
        $60.49$                                        & 
        $0.97$                                         & 
        Fig.~\ref{fig:stage_0_execution}, red line     \\
        1                                              & 
        $T^{UC1}_{2}$                                  & 
        $5.21\times 10^{-2}$                           & 
        $128.53$                                       & 
        $0.96$                                         & 
        Fig.~\ref{fig:stage_1_execution}, green line   \\
        1                                              & 
        $T^{UC2}_{1}$                                  & 
        $0.93$                                         &
        $2.45$                                         &
        $0.97$                                         &
        Fig.~\ref{fig:geo_stage_0_execution}, red line \\
        1                                              &
        $T^{UC2}_{2}$                                  &
        $5.21\times 10^{-2}$                           &
        $128.53$                                       &
        $0.96$                                         &
        Fig.~\ref{fig:geo_stage_1_execution}, green line   \\
        2                                              &
        $T^{UC1}_{1}$                                  &
        $3.17\times 10^{-2}$                           &
        $64.81$                                        &
        $0.92$                                         &
        Fig.~\ref{fig:stage_1_execution_des2}, red line     \\
        2                                              &
        $T^{UC1}_{2}$                                  &
        $4.71\times 10^{-2}$                           &
        $95.83$                                        &
        $0.95$                                         &
        Fig.~\ref{fig:stage_2_execution_des2}, green line   \\
        2                                              &
        $T^{UC2}_{1}$                                  &
        $0.62$                                         &
        $1.52$                                         &
        $0.61$                                         &
        Fig.~\ref{fig:geo_stage_0_execution_des2}, red line     \\
        2                                              &
        $T^{UC2}_{2}$                                  &
        $3.16 \times 10^{-2}$                                         &
        $0.29$                                         &
        $0.51$                                         &
        Fig.~\ref{fig:geo_stage_1_execution_des2}, green line   \\
        2.A                                            &
        $T^{UC1}_{1}$                                  &
        $2.74\times 10^{-2}$                           &
        $49.03$                                        &
        $0.94$                                         &
        N/A     \\
        2.A                                            &
        $T^{UC1}_{2}$                                  &
        $4.80\times 10^{-2}$                           &
        $87.60$                                        &
        $0.95$                                         &
        N/A   \\
        2.A                                            &
        $T^{UC2}_{1}$                                  &
        $0.54$                                         &
        $1.51$                                         &
        $0.76$                                         &
        N/A     \\
        2.A                                            &
        $T^{UC2}_{2}$                                &
        $2.82 \times 10^{-2}$                                         &
        $0.26$                                         &
        $0.89$                                         &
        N/A     \\
		\bottomrule
    \end{tabular}
    \caption{Fitted parameter values of Eq.~\ref{eq:des1_til} using a
             non-linear least squares algorithm to fit our experimental
             data.}\label{tab:fit_par_val}
\end{table*}

Fig.~\ref{fig:stage_1_execution} shows the execution time of $T^{UC1}_2$ as a
function of the image size. This task presents a different behavior than
$T^{UC1}_{1}$, as the code executed is different. Note the slightly stronger
positive skew of the data compared to that of
Fig.~\ref{fig:stage_0_execution} but still consistent with our conclusion
that deviations are mostly due to fluctuations in the node performance (i.e.,
different code similar fluctuations).

Similar to $T^{UC1}_{1}$, Fig.~\ref{fig:stage_1_execution} shows a weak
correlation between the execution time of $T^{UC1}_{2}$ and image size. In
addition, the variance per bin is relatively similar across bins, as expected
based on the analysis of $T^{UC1}_{1}$. The box-plot and mean execution
time indicate that a linear function is a good candidate for a model of
$T^{UC1}_{2}$. We fitted a linear function, as in Eq.~\ref{eq:des1_til}, to
the execution time as a function of the image size for the same bins as
$T^{UC1}_{1}$.

Fig.~\ref{fig:geo_stage_1_execution} shows the execution time of
$T^{UC2}_{2}$ as function of the size of the image pair. We notice a weak
correlation between $T^{UC2}_{2}$ execution time and the size of the image
pair. Further, we see a similar variance among the bins as that measured for
$T^{UC1}_{2}$. We notice that the execution time becomes positively skewed as
the image pair size increases, as shown in
Figs~\ref{fig:geo_stage_0_execution} and~\ref{fig:geo_stage_1_execution}.
Node usage increases with image size, making the observed fluctuations
consistent with the analysis from UC1.

Based on Table~\ref{tab:fit_par_val}, the $R^2$ values for $T^{UC1}_{1}$,
$T^{UC1}_{2}$, $T^{UC2}_{1}$ and $T^{UC2}_{2}$ show a good fit of the
respective lines to the actual data. As a result, we can conclude that our
estimated functions are validated.

\subsection{Experiment~1, Design~2:}\label{ssec:des2analysis}

\begin{figure*}
    \centering
    \begin{subfigure}[b]{0.49\textwidth}
        \includegraphics[width=\linewidth]{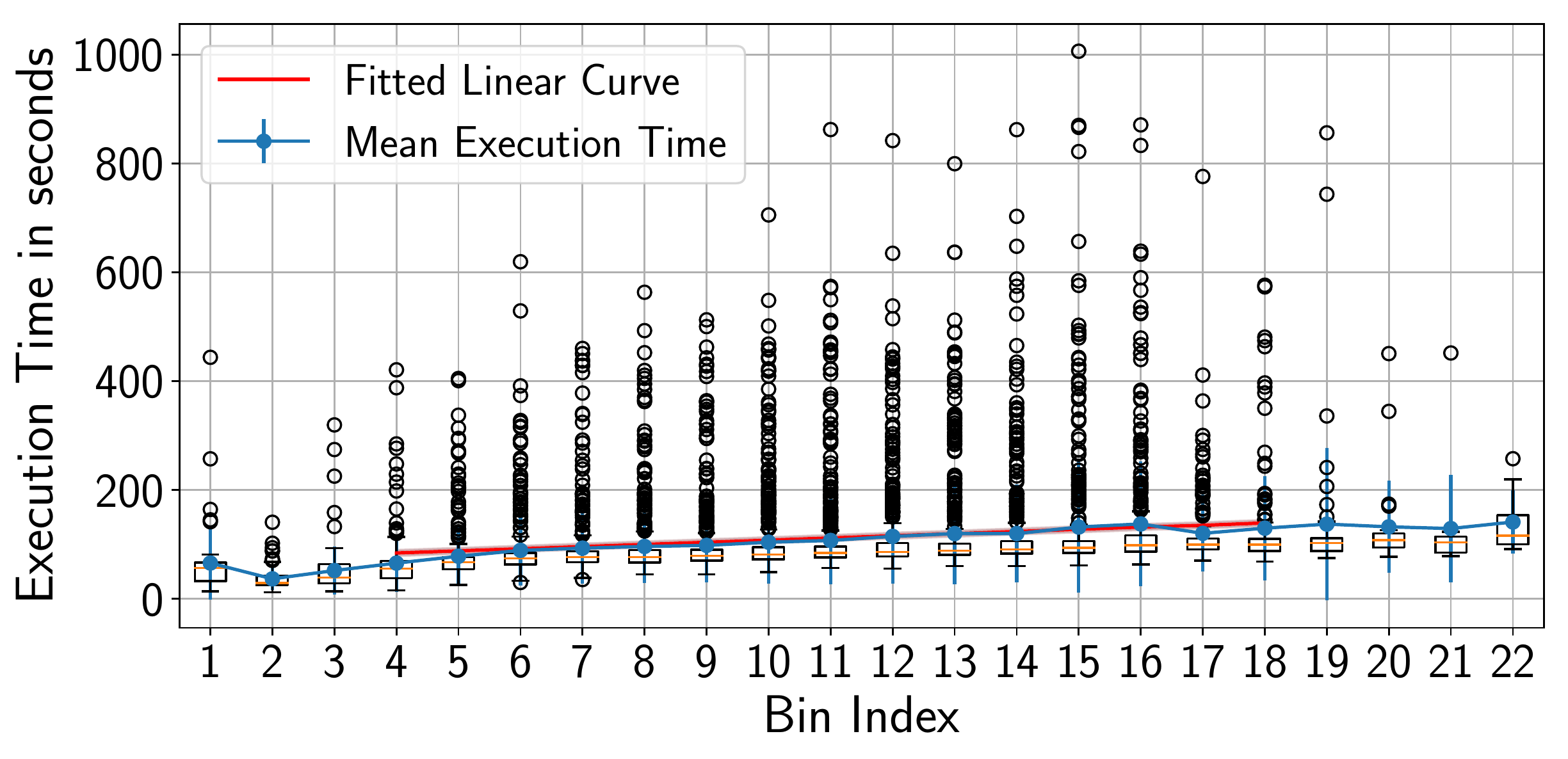}
        \caption{}\label{fig:stage_1_execution_des2}
    \end{subfigure}%
    ~ 
    \begin{subfigure}[b]{0.49\textwidth}
        \includegraphics[width=\linewidth]{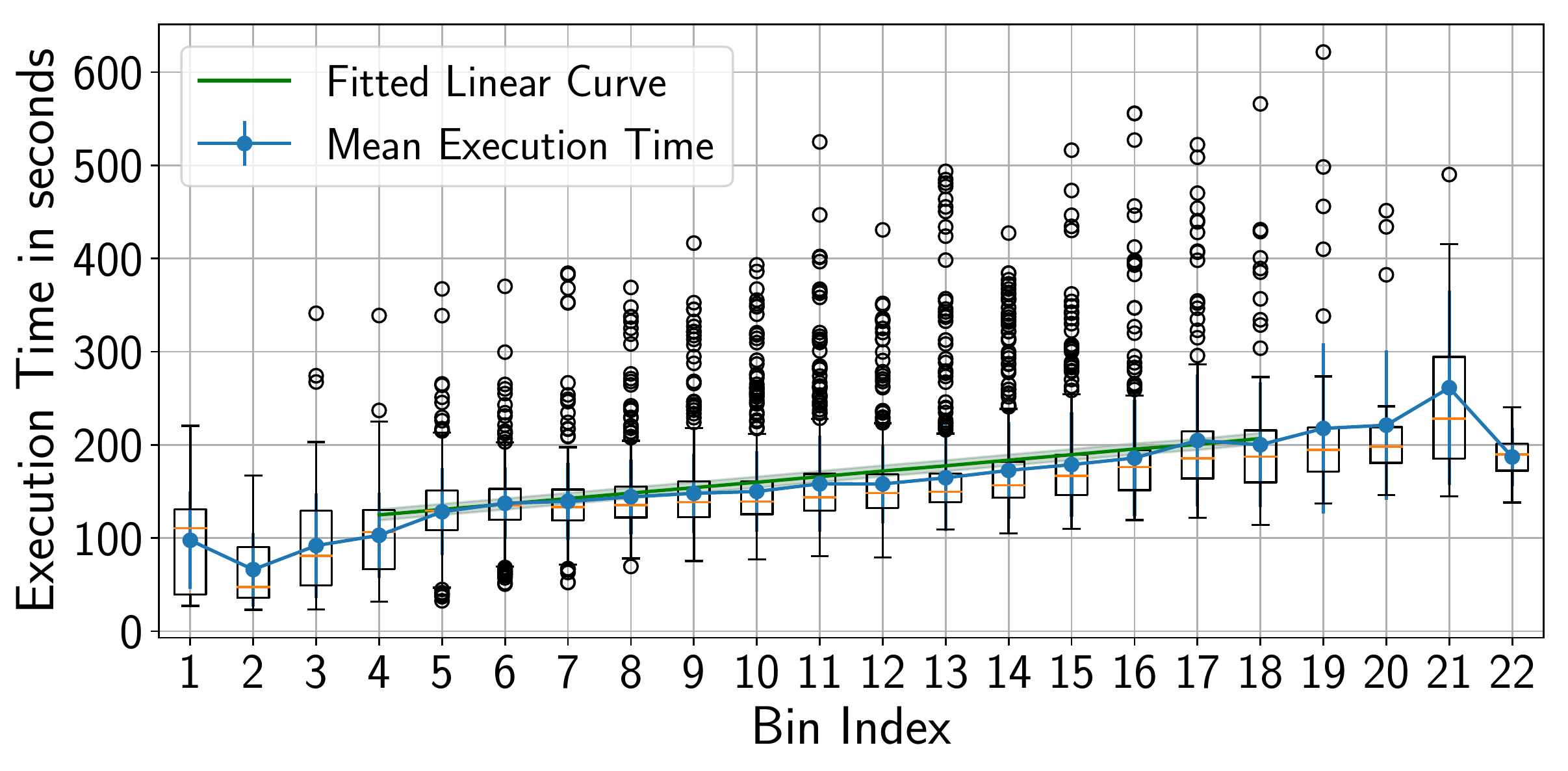}
        \caption{}\label{fig:stage_2_execution_des2}
    \end{subfigure}
    \caption{Experiment~1, Design~2, UC1: Box-plots of (a) $T^{UC1}_{1}$ and
    (b) $T^{UC1}_{2}$ execution times, means and standard deviations ($STDs$)
    for $125$~MB image pair size bins. Red line shows fitted linear function
    for $T^{UC1}_{1}$, green line for $T^{UC1}_{2}$. Red shadow shows
    confidence interval for $T^{UC1}_{1}$, green shadow for
    $T^{UC1}_{2}$.}\label{fig:execution_des2}
\end{figure*}

Fig.~\ref{fig:stage_1_execution_des2} shows the execution time of
$T^{UC1}_{1}$ as a function of the image size for Design~2. In principle,
design differences in middleware that execute tasks as independent programs
should not directly affect task execution time. In this type of middleware,
task code is independent from that of the middleware: once tasks execute, the
middleware waits for each task to return. Nonetheless, in real scenarios with
concurrency and heterogeneous tasks, the middleware may perform operations on
multiple tasks while waiting for others to return. Accordingly, in Design~2
we observe an execution time variation comparable to that observed with
Design~1 but Fig.~\ref{fig:stage_1_execution_des2} shows a stronger positive
skew of the data for Design~2 than Fig.~\ref{fig:stage_0_execution} for
Design~1.

We investigated the positive skew of the data observed in
Fig.~\ref{fig:stage_1_execution_des2} by comparing the system load of a
compute node when executing the same number of tiling tasks for Design~1 and
2. The system load of Design~2 was higher than that of Design~1. 
As we used the same type of task, image and task concurrency, we conclude
that the middleware implementing Design~2 uses more compute resources than
that used for Design~1. Due to concurrency, the middleware of Design~2
competes for resources with the tasks, momentarily slowing down their
execution. This is consistent with the architectural differences across the
two designs: Design~2 requires resources to manage queues and data movement
while Design~1 has only to schedule and launch tasks on each node.

Design~2 also produces a much stronger positive skew of $T^{UC1}_{2}$
execution time compared to executing $T^{UC1}_{2}$ with Design~1 (see
Fig.~\ref{fig:stage_2_execution_des2}). $T^{UC1}_{2}$ executes on GPU and
$T^{UC1}_{1}$ on CPU but their execution times have comparable skew in
Design~2. This further supports our hypothesis that the long tail of the
distribution of $T^{UC1}_{1}$ and especially $T^{UC1}_{2}$ execution times
depends on the competition for main memory and I/O between the middleware and
the executing tasks.

Table~\ref{tab:fit_par_val} shows the model parameters for both tasks and
their respective $R^2$ values. $R^2$ are worse compared to Design~1. This is
expected based on the positive skew of the data observed in Design~2.

\begin{figure*}
    \centering
    \begin{subfigure}[b]{0.49\textwidth}
        \includegraphics[width=\linewidth]{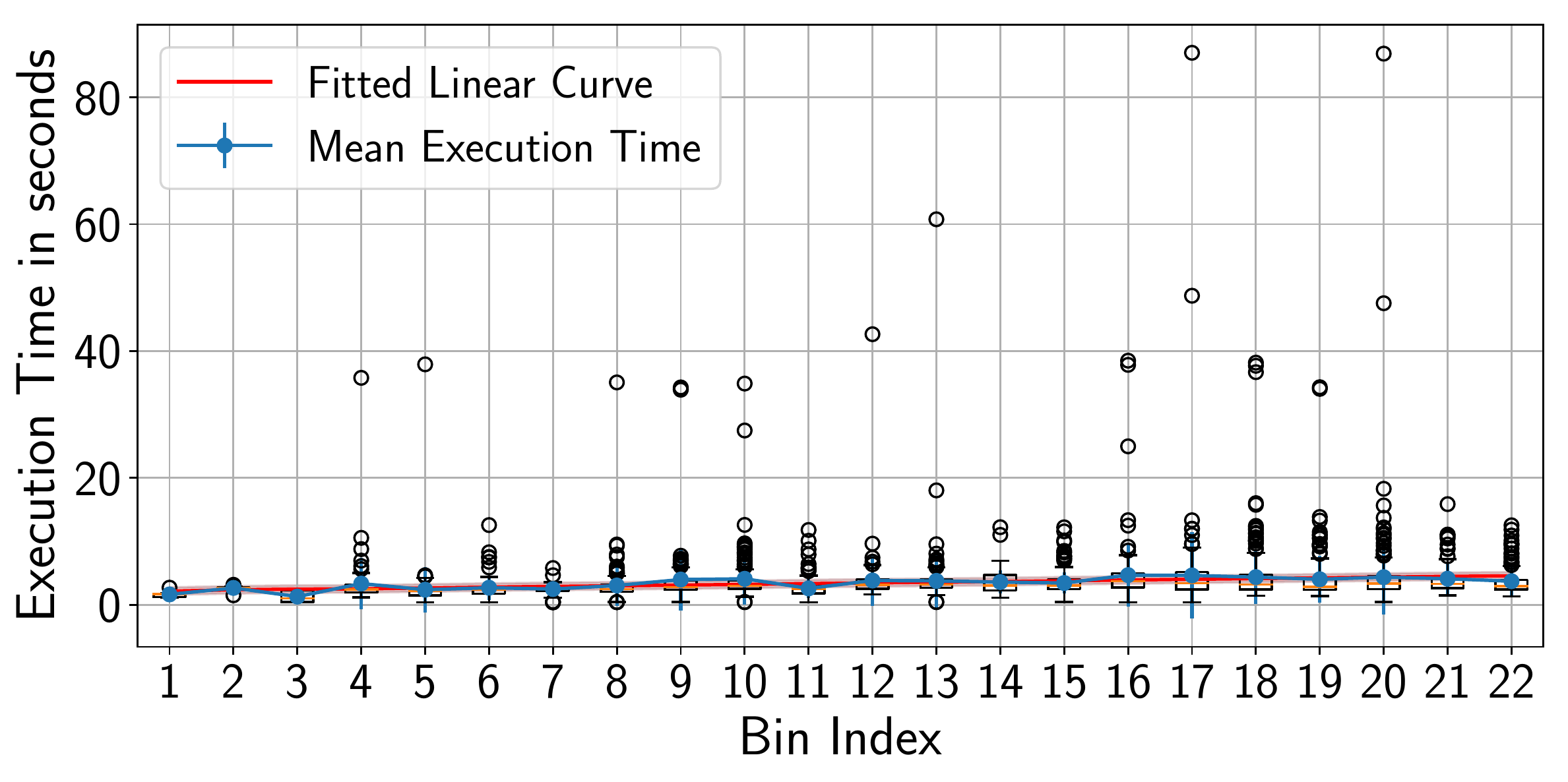}
        \caption{}\label{fig:geo_stage_0_execution_des2}
    \end{subfigure}%
    ~ 
    \begin{subfigure}[b]{0.49\textwidth}
        \includegraphics[width=\linewidth]{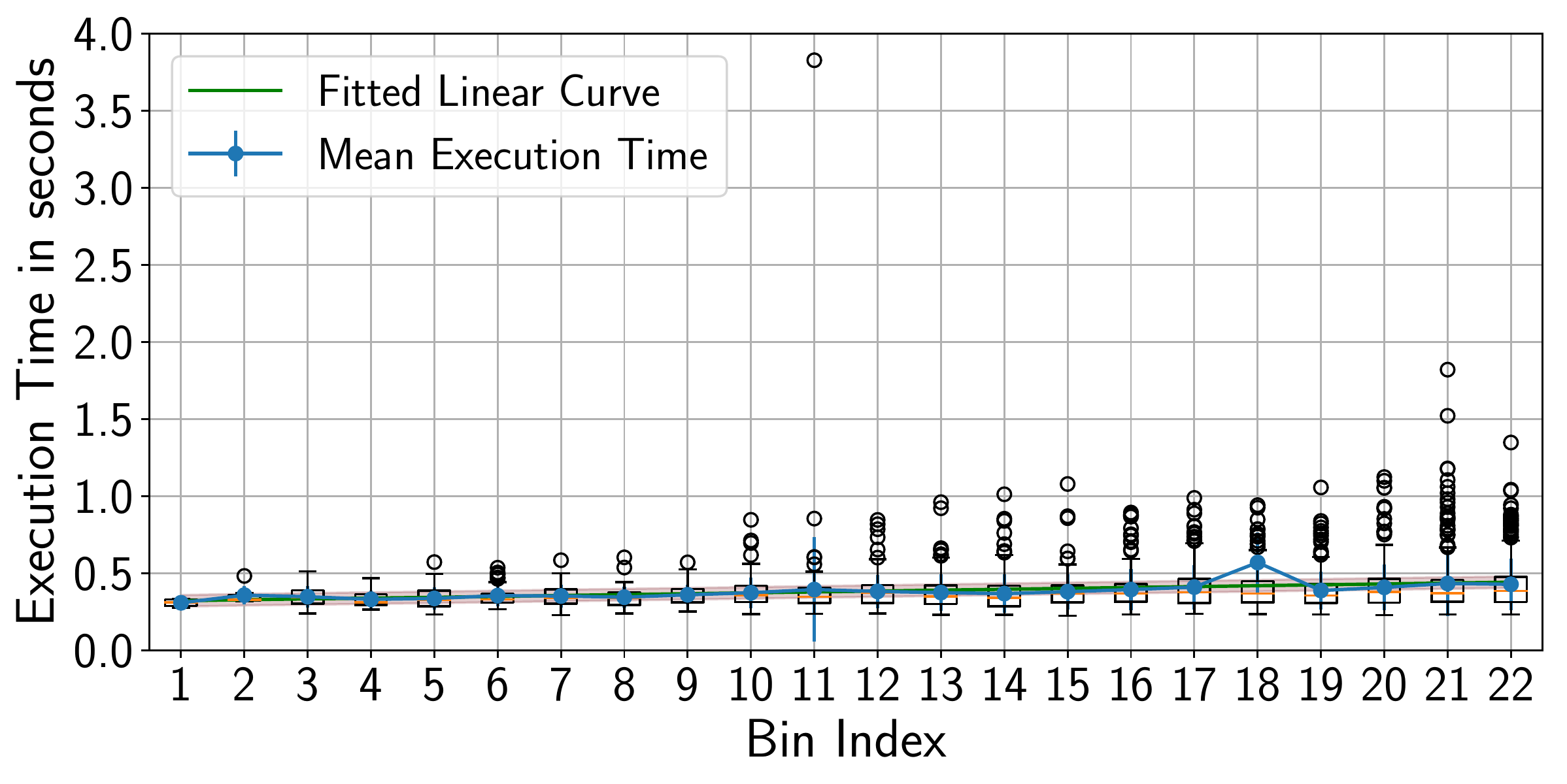}
        \caption{}\label{fig:geo_stage_1_execution_des2}
    \end{subfigure}
    \caption{Experiment~1, Design~2, UC2: Box-plots of
    (a) $T^{UC2}_{1}$ and (b) $T^{UC2}_{2}$ execution times, means and
    standard deviations ($STDs$) for $187$~KB image size bins. Red line shows
    fitted linear function for $T^{UC2}_{1}$, green line for $T^{UC2}_{2}$.
    Red shadow shows confidence interval for $T^{UC2}_{1}$, green shadow for
    $T^{UC2}_{2}$.}\label{fig:geo_execution_des2}
\end{figure*}

Fig.~\ref{fig:geo_stage_0_execution_des2} shows the execution time of
$T^{UC2}_{1}$ as a function of image pair size for Design~2. We notice an
execution time variation comparable to the one shown in
Fig.~\ref{fig:geo_stage_0_execution} for Design~1. Nonetheless, while the
number of outliers is lower than in Design~1, their spread is higher: between
10sec and 80sec for Design~2 compared to 5sec and 16sec for Design~1. As with
UC1, this positive skewness is due to the increased system load per compute
node with Design~2. The fitted parameter values are shown in
Table~\ref{tab:fit_par_val} indicating a good fit.

Fig.~\ref{fig:geo_stage_1_execution_des2} shows the execution time of
$T^{UC2}_{2}$. We notice a skew of the data similar to what observed for
$T^{UC2}_{2}$ in Design~1 and a larger spread of the outliers. The latter
supports our hypothesis that the execution time of the task is sensitive to
resource competition, especially related to the use of the file system. This
wider spread supports a worse fit of our model, as shown in
Table~\ref{tab:fit_par_val}. Spread apart, there is not much difference
between the execution times of $T^{UC2}_{2}$ tasks in Design~1 and Design~2.

\subsection{Experiment~1, Design~2.A:}

Similarly to the analysis for Design~1 and 2, we fitted UC1 and UC2 data from
Design~2.A to Eq.~\ref{eq:des1_til}. The fitted parameter are shown in
Table~\ref{tab:fit_par_val}. Based on $R^2$, we can conclude that all model
are good fits for their respective data.

The results of experiment 1 indicate that with Design~2.A, on average, there
is a decrease in the execution time of $T_{1}$ and an increase in that of
$T{2}$ compared to Design~2, for both use cases. Design~2.A requires one
queue more than Design~2 for $T^{UC1}_{1}$ and therefore more resources for
its implementation. This can explain the slowing of $T_{2}$ but not the
speedup of $T_{1}$. This requires further investigation, measuring whether
the performance fluctuations of compute nodes are larger than measured so
far.

\begin{figure*}[ht!]
    \centering
    \begin{subfigure}[b]{0.49\textwidth}
        \includegraphics[width=\linewidth]{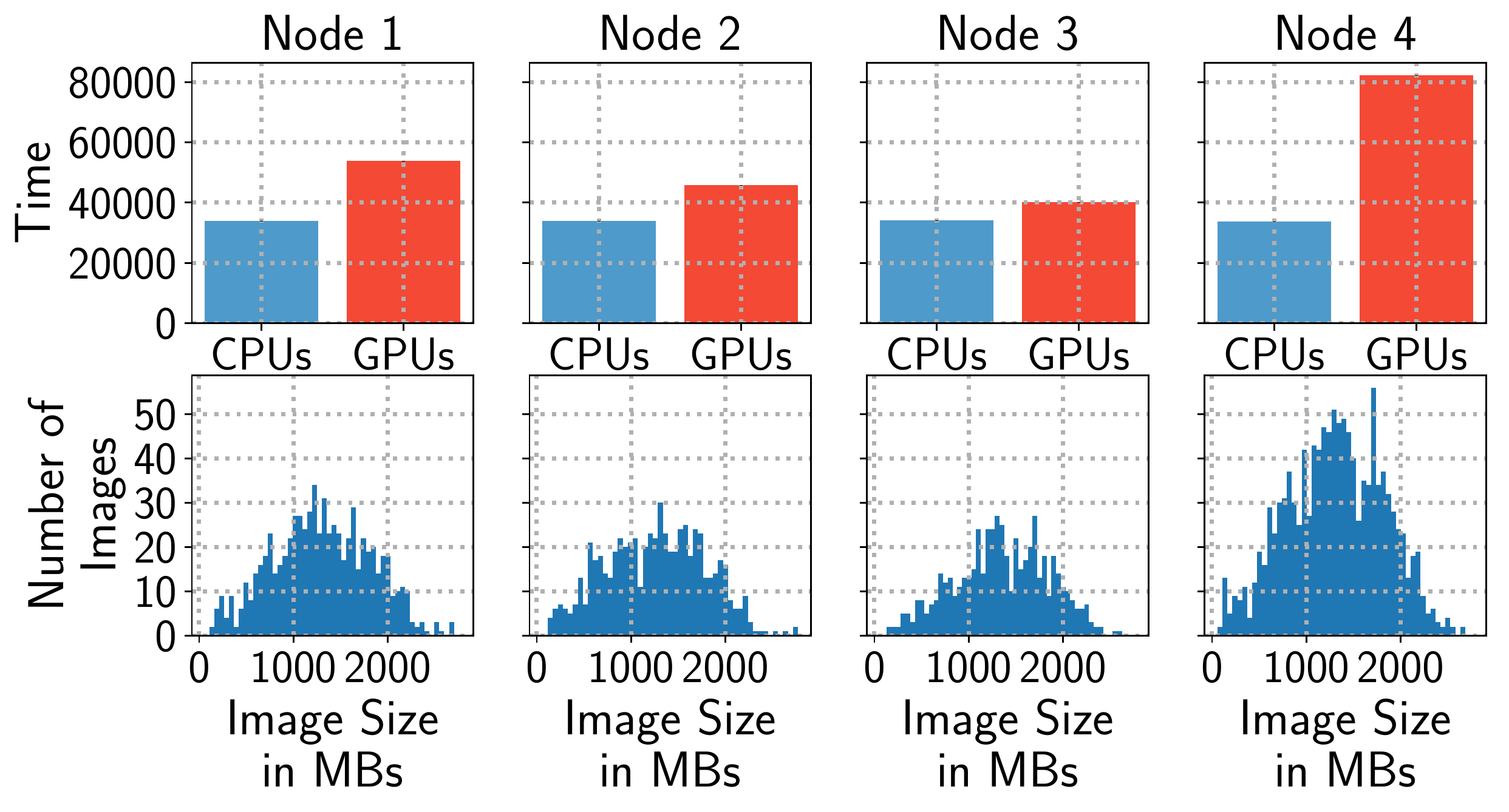}
        \caption{}
        \label{fig:design2_timeline}
    \end{subfigure}%
    ~ 
    \begin{subfigure}[b]{0.49\textwidth}
        \includegraphics[width=\linewidth]{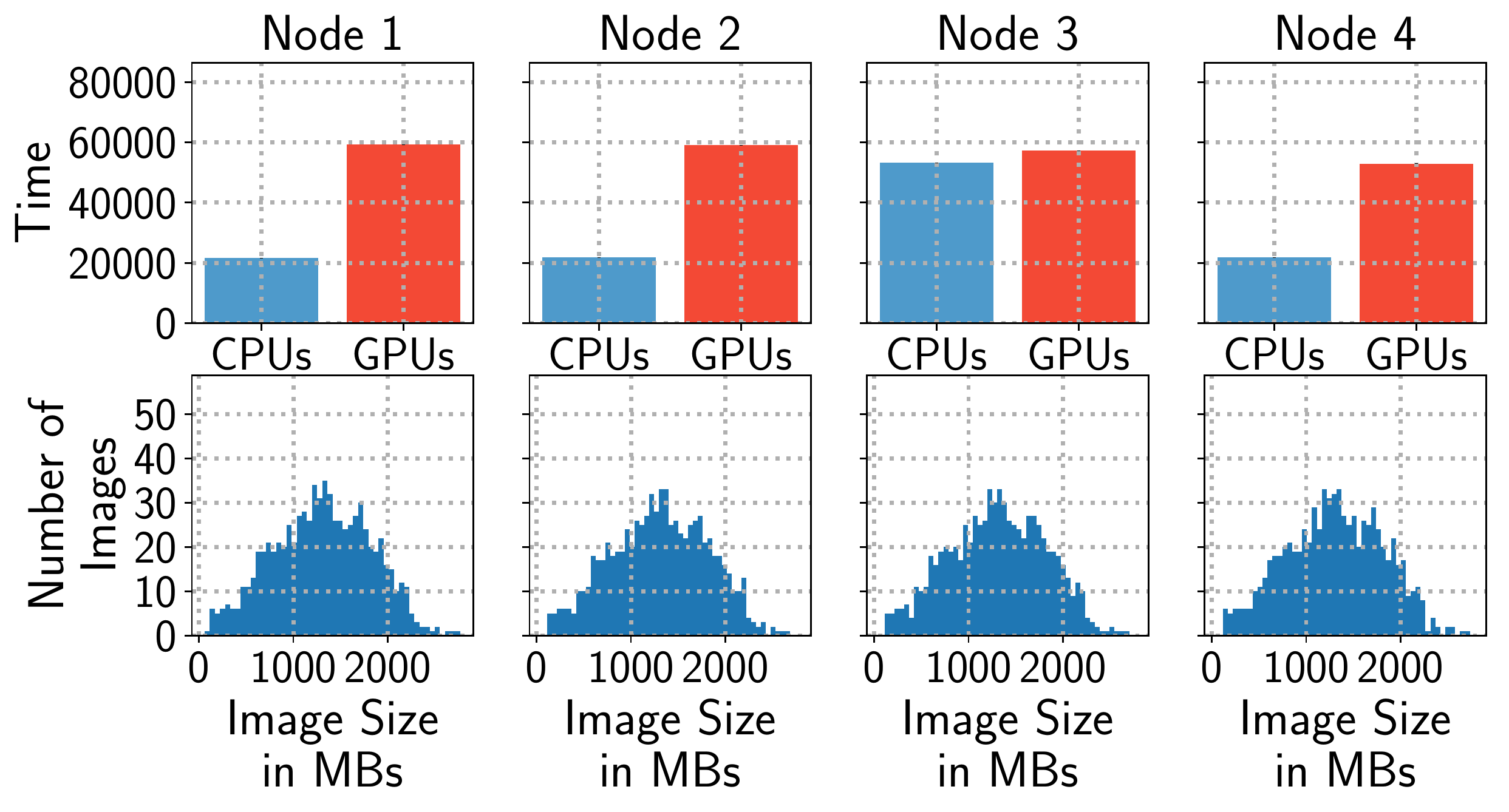}
        \caption{}
        \label{fig:design2a_timeline}
    \end{subfigure}
    \caption{Experiment~1, Design~2.A, UC1 execution time of $T^{UC1}_{1}$
    (blue) and $T^{UC1}_{2}$ (red), and distributions of image size per node
    for (a) Design~2 and (b) Design~2.A.}\label{fig:design_balancing}
\end{figure*}

\begin{figure*}[ht!]
	\centering
	\begin{subfigure}[b]{0.49\textwidth}
		\includegraphics[width=\linewidth]{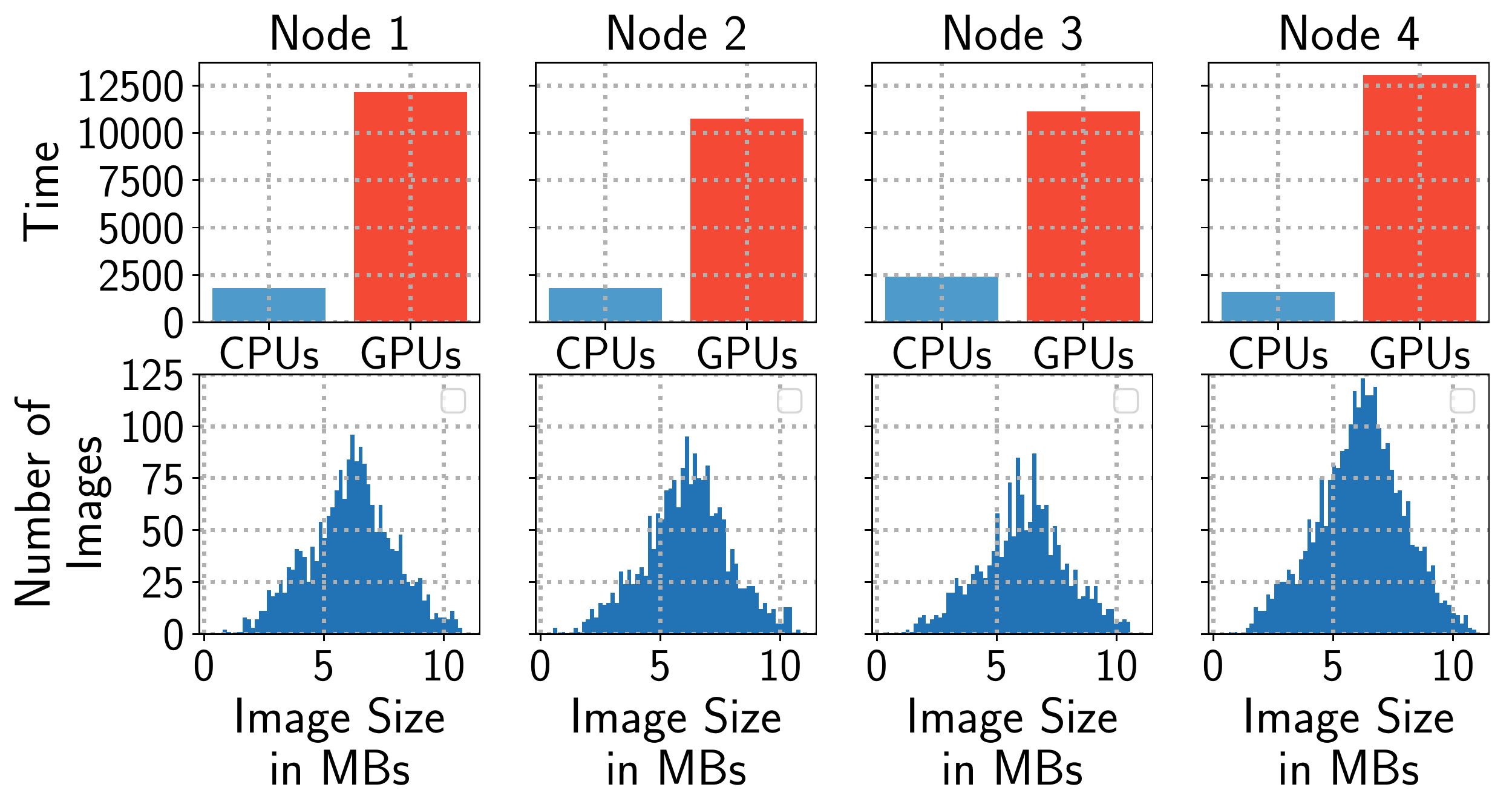}
		\caption{}
		\label{fig:geo_design2_timeline}
	\end{subfigure}%
	~ 
	\begin{subfigure}[b]{0.49\textwidth}
		\includegraphics[width=\linewidth]{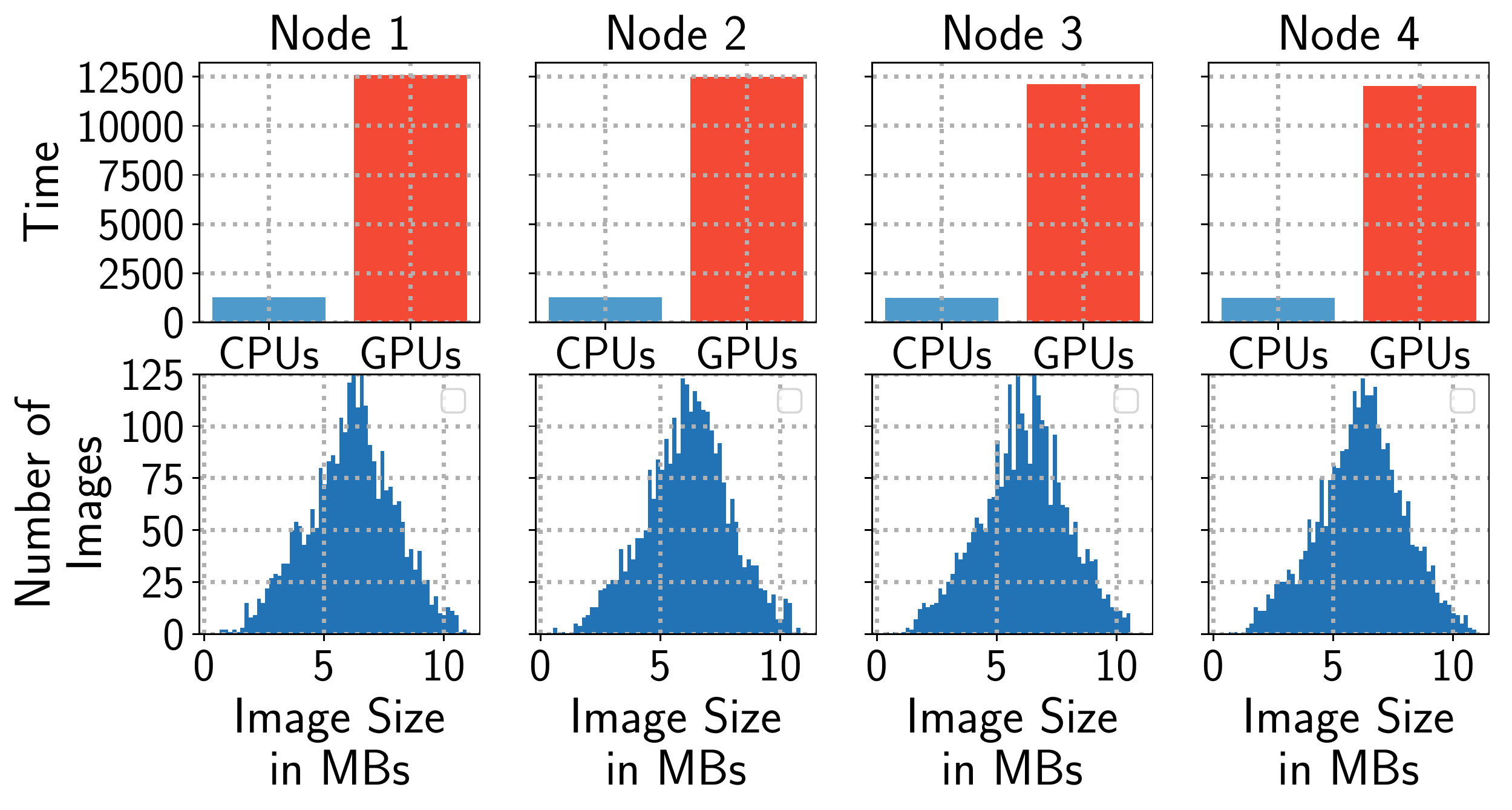}
		\caption{}
		\label{fig:geo_design2a_timeline}
	\end{subfigure}
    \caption{Experiment~1, Design~2.A, UC2 execution time of $T^{UC2}_{1}$
    (blue) and $T^{UC2}_{2}$ (red), and distributions of image size per node
    for (a) Design~2 and (b) Design~2.A.}\label{fig:geo_design_balancing}
\end{figure*}

As discussed in~\S\ref{ssec:approach2}, balancing of workflow execution
differs between Design~2 and Design~2.A. Figs.~\ref{fig:design2_timeline}
and~\ref{fig:geo_design2_timeline} show that the task $T_{1}$ of the two use
cases can work on a different number of images but all $T_{1}$ tasks
concurrently execute for a similar duration. The histograms in
Figs.~\ref{fig:design2_timeline} and~\ref{fig:geo_design2_timeline} also show
that this balancing can result in different input distributions for each
compute node, affecting the total execution time of the $T_{2}$ tasks on each
node. Thus, Design~2 can create imbalances in the time to completion of
$T_{2}$, as shown by the red bars in both Figs.~\ref{fig:design2_timeline}
and~\ref{fig:geo_design2_timeline}.

Design~2.A addresses these imbalances by early binding images to compute
nodes. Comparing the lower part of Fig.~\ref{fig:design2_timeline} with
Fig.~\ref{fig:design2a_timeline} and Fig.~\ref{fig:geo_design2_timeline} with
Fig.~\ref{fig:geo_design2a_timeline}, we notice the difference between the
distributions of image size for each node between Design~2 and 2.A. In
Design~2.A, due to the modeled correlation between time to completion and the
size of the processed image, the similar distribution of the size of the
images bound to each compute node balances the total processing time of the
workflow across multiple nodes.

Note that Figs.~\ref{fig:design_balancing} and~\ref{fig:geo_design_balancing}
show just one of the runs we perform for this experiment. Due to the random
pulling of images from a global queue performed by Design~2, each run shows
different distributions of image sizes across nodes, leading to large
variations in the total execution time of the workflow.

Fig.~\ref{fig:design2a_timeline} shows also an abnormal behavior of one
compute node: For images larger than 1.5~GBs, Node 3 CPU performance is
markedly slower than other nodes when executing $T^{UC1}_{1}$. Different from
Design~2, Design~2.A can balance these fluctuations in $T^{UC1}_{1}$ as far
as they don't starve $T^{UC1}_{2}$ tasks.

Fig.~\ref{fig:geo_design2a_timeline} shows a more balanced and decreasing
execution time of $T^{UC2}_{1}$ among the 4 nodes, compared to Design~2. We
investigated the decreasing of the execution time and we explained it with
the different input distribution of the total size of the image pairs
(smaller) in Node 3 and 4 compared to Node 1 and Node 2. This image size
variation can create a significant fluctuations in the execution time.

\subsection{Experiment~2: Resource Utilization}\label{ssec:exp2}

Resource utilization varies across Design~1, 2 and 2.A. In Design~1, the
runtime system (RTS), i.e, RADICAL-Pilot, is responsible for scheduling and
executing tasks. For UC1, $T^{UC1}_{1}$ is memory intensive and, as a
consequence, we were able to concurrently execute 3 $T^{UC1}_{1}$ on each
compute node, using only 3 of the 32 available CPU cores. We were instead
able to execute 2 $T^{UC1}_{2}$ concurrently on each node, using all the
available GPUs. Assuming ideal concurrency among the 4 compute nodes we
utilized in our experiments, the theoretical maximum utilization per node
would be 10.6\% for CPUs and 100\% for GPUs.

For UC2, $T^{UC2}_{1}$ uses GPU-SIFT to process and match a pair of images.
$T^{UC2}_{1}$ is a GPU-memory intensive task that requires an amount of
memory proportional to the size of the image pair on which to perform image
matching. $T^{UC2}_{2}$ runs on CPU and requires only one core to perform the
RANSAC filtration. We were able to execute 2 $T^{UC2}_{1}$ and up to 2
$T^{UC2}_{2}$ concurrently on every compute node, utilizing all of the 8 GPUs
but only up to 2 of the 32 available cores per node.

\begin{figure*}
    \centering
    \begin{subfigure}[b]{0.33\textwidth}
        \includegraphics[width=\linewidth]{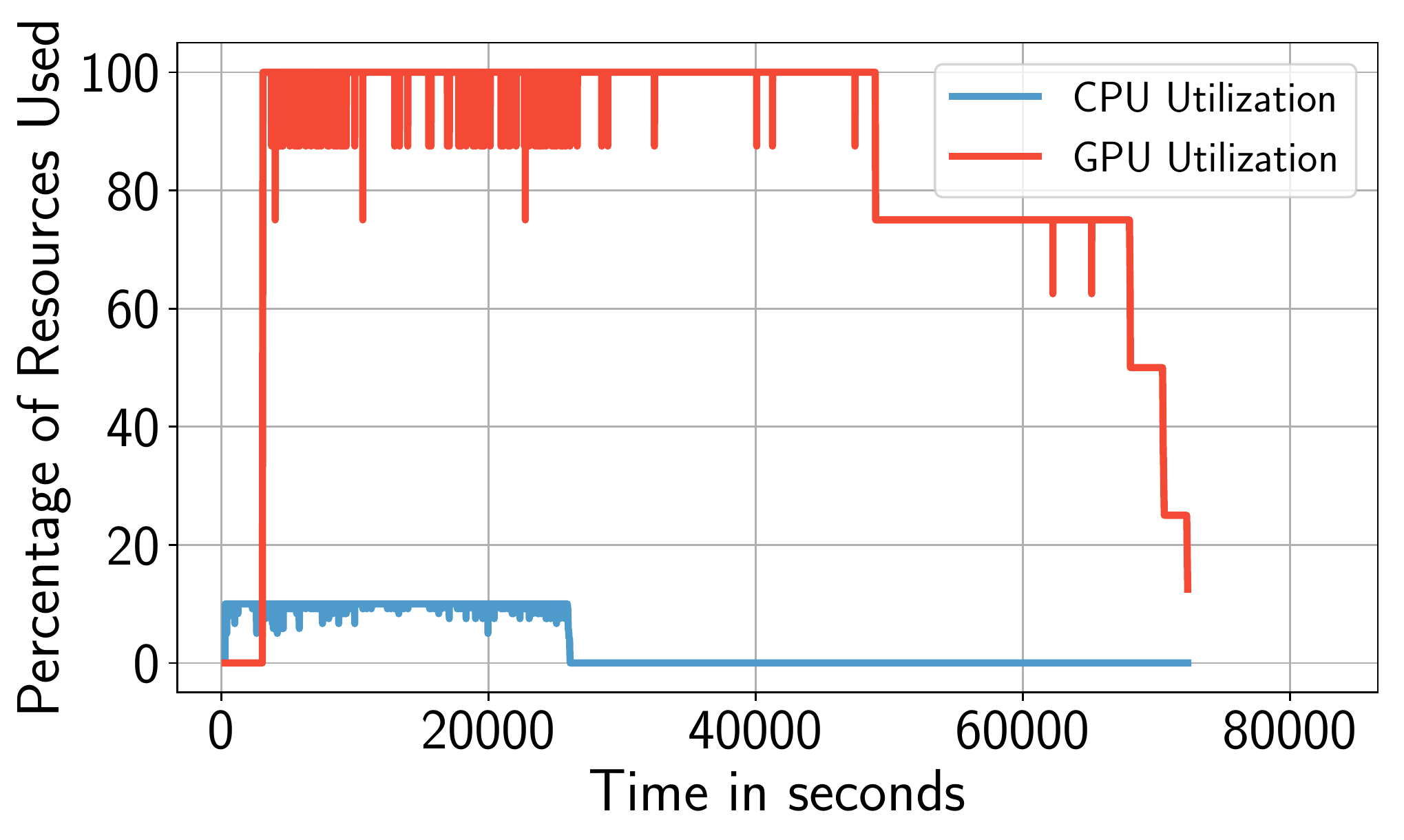}
        \caption{}
        \label{fig:design1util}
    \end{subfigure}%
    ~ 
    \begin{subfigure}[b]{0.33\textwidth}
        \includegraphics[width=\linewidth]{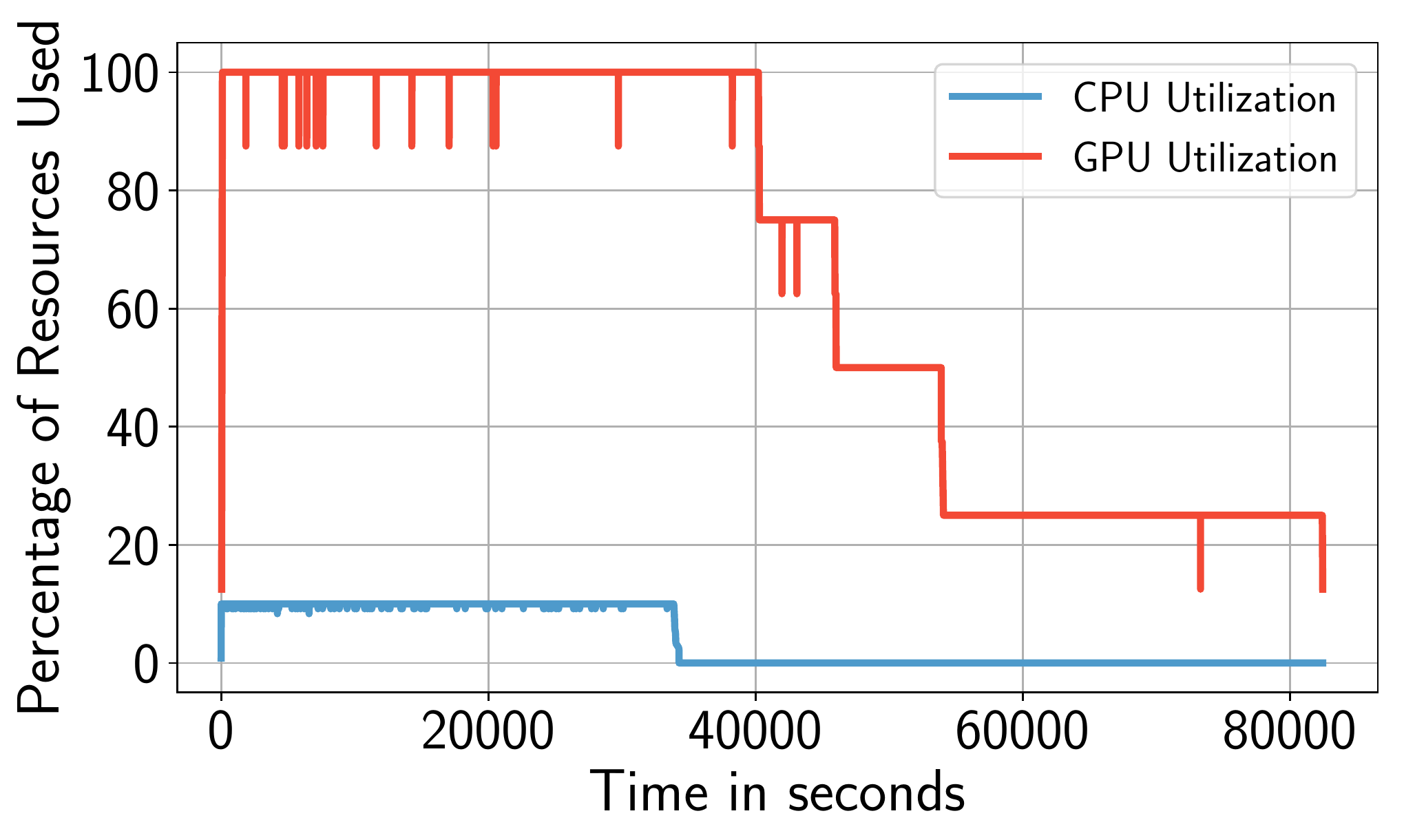}
        \caption{}
        \label{fig:design2util}
    \end{subfigure}%
    ~ 
    \begin{subfigure}[b]{0.33\textwidth}
        \includegraphics[width=\linewidth]{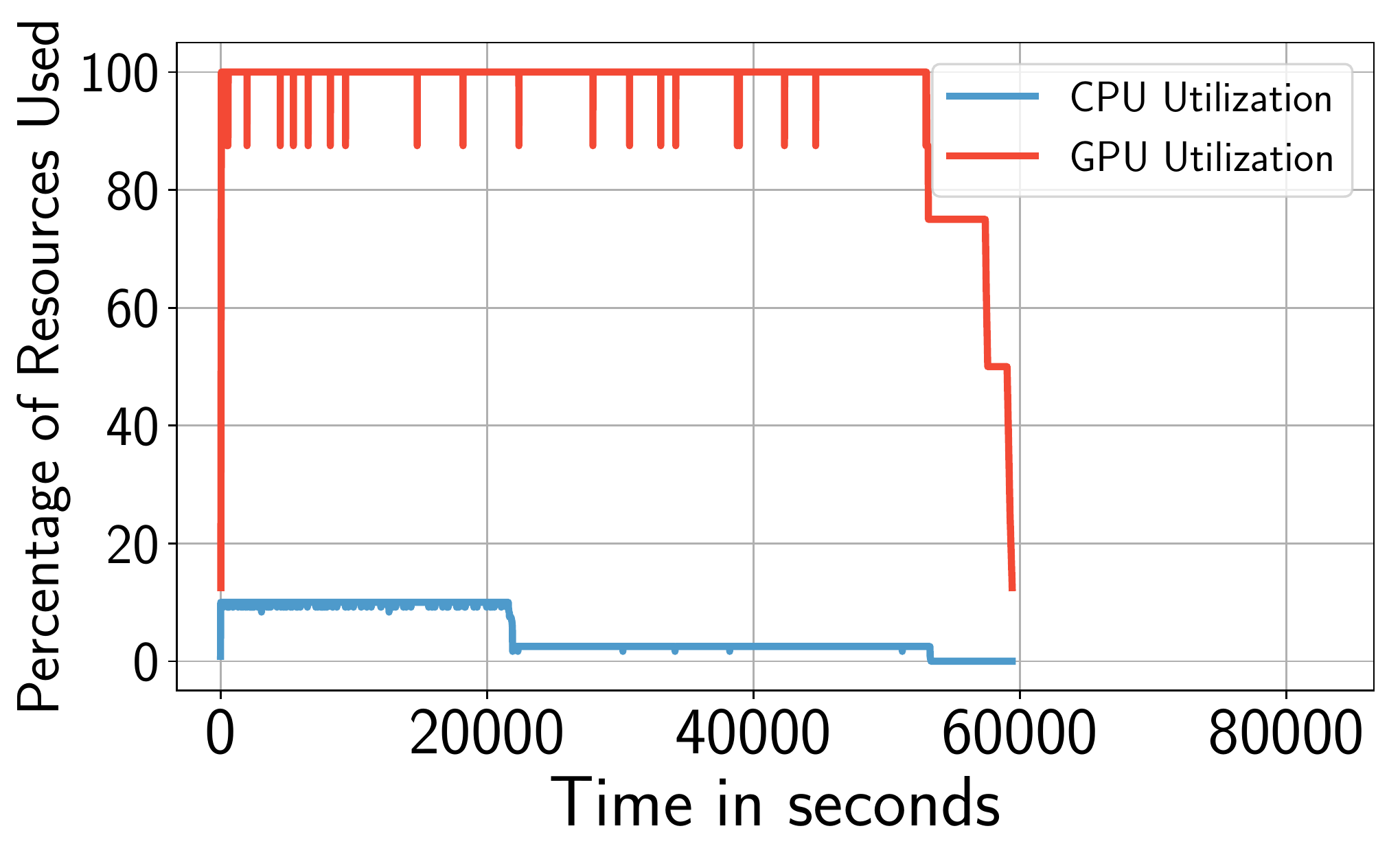}
        \caption{}
        \label{fig:design2autil}
    \end{subfigure}
    \caption{Experiment~2, UC1 Percentage of CPU and GPU utilization for: (a)
    Design~1; (b) Design~2, and (3) Design~2.A.}\label{fig:Utilization}
\end{figure*}

\begin{figure*}
    \centering
    \begin{subfigure}[b]{0.33\textwidth}
        \includegraphics[width=\linewidth]{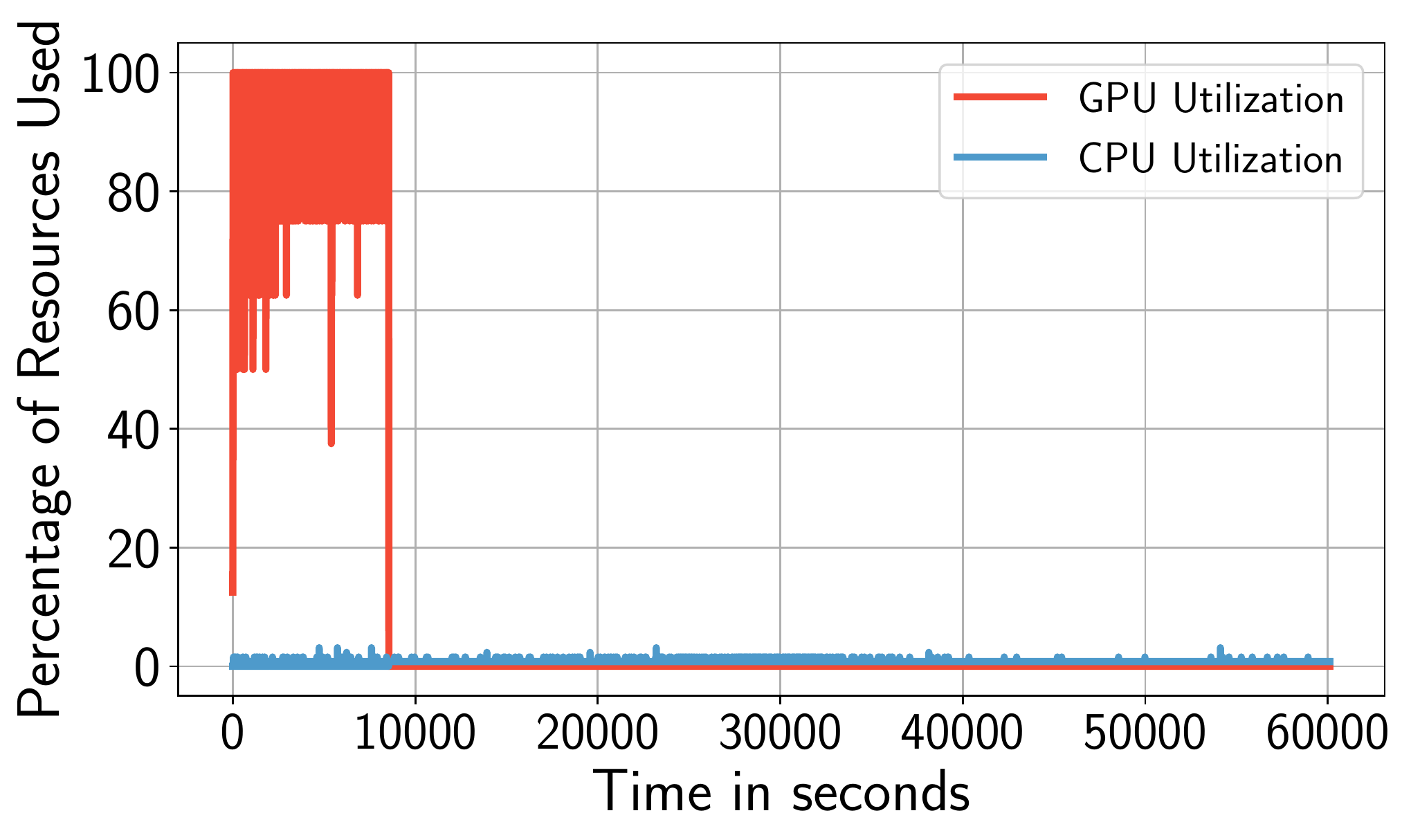}
        \caption{}
        \label{fig:geo_design1util}
    \end{subfigure}%
    ~ 
    \begin{subfigure}[b]{0.33\textwidth}
        \includegraphics[width=\linewidth]{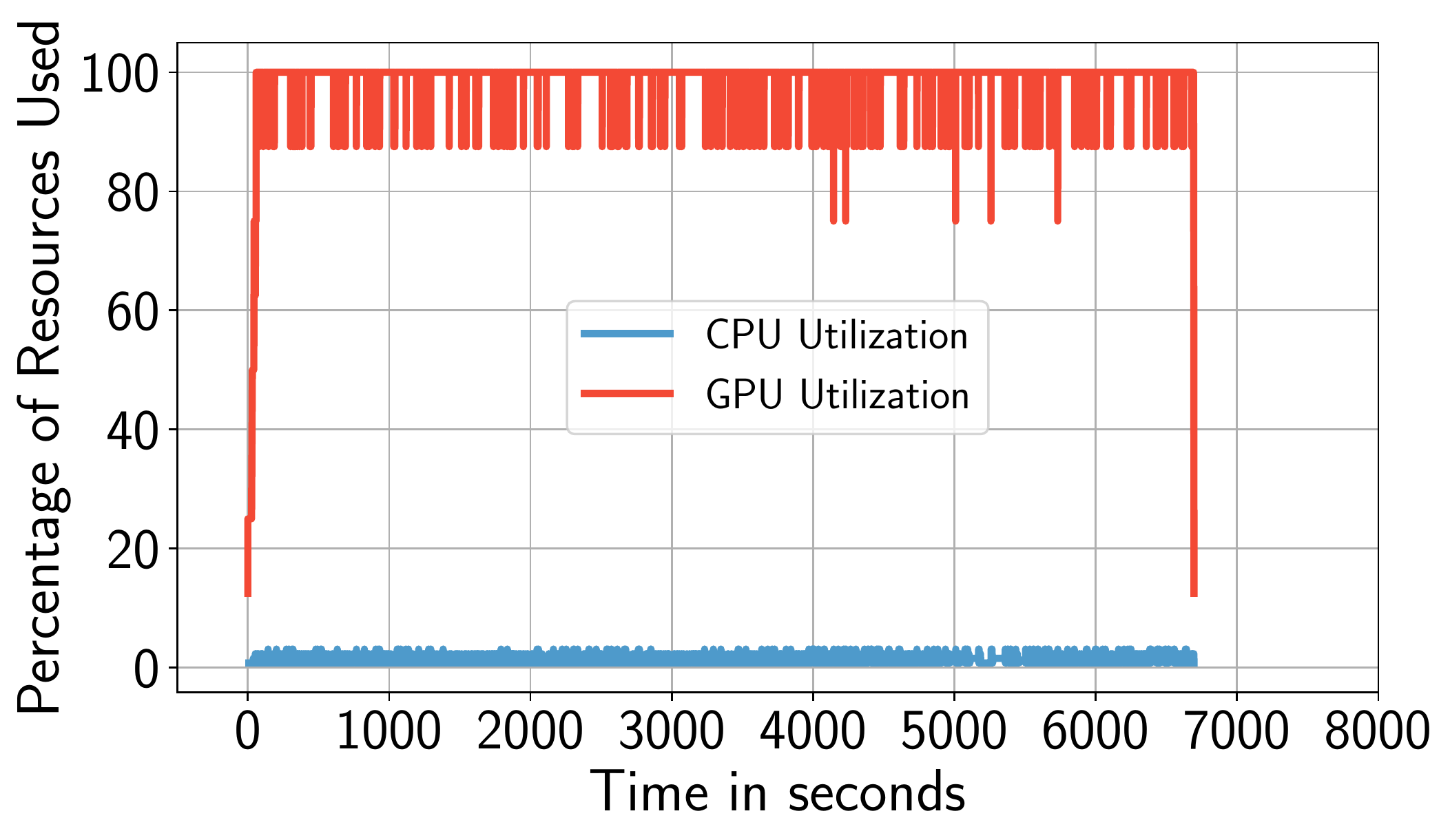}
        \caption{}
        \label{fig:geo_design2util}
    \end{subfigure}%
    ~ 
    \begin{subfigure}[b]{0.33\textwidth}
        \includegraphics[width=\linewidth]{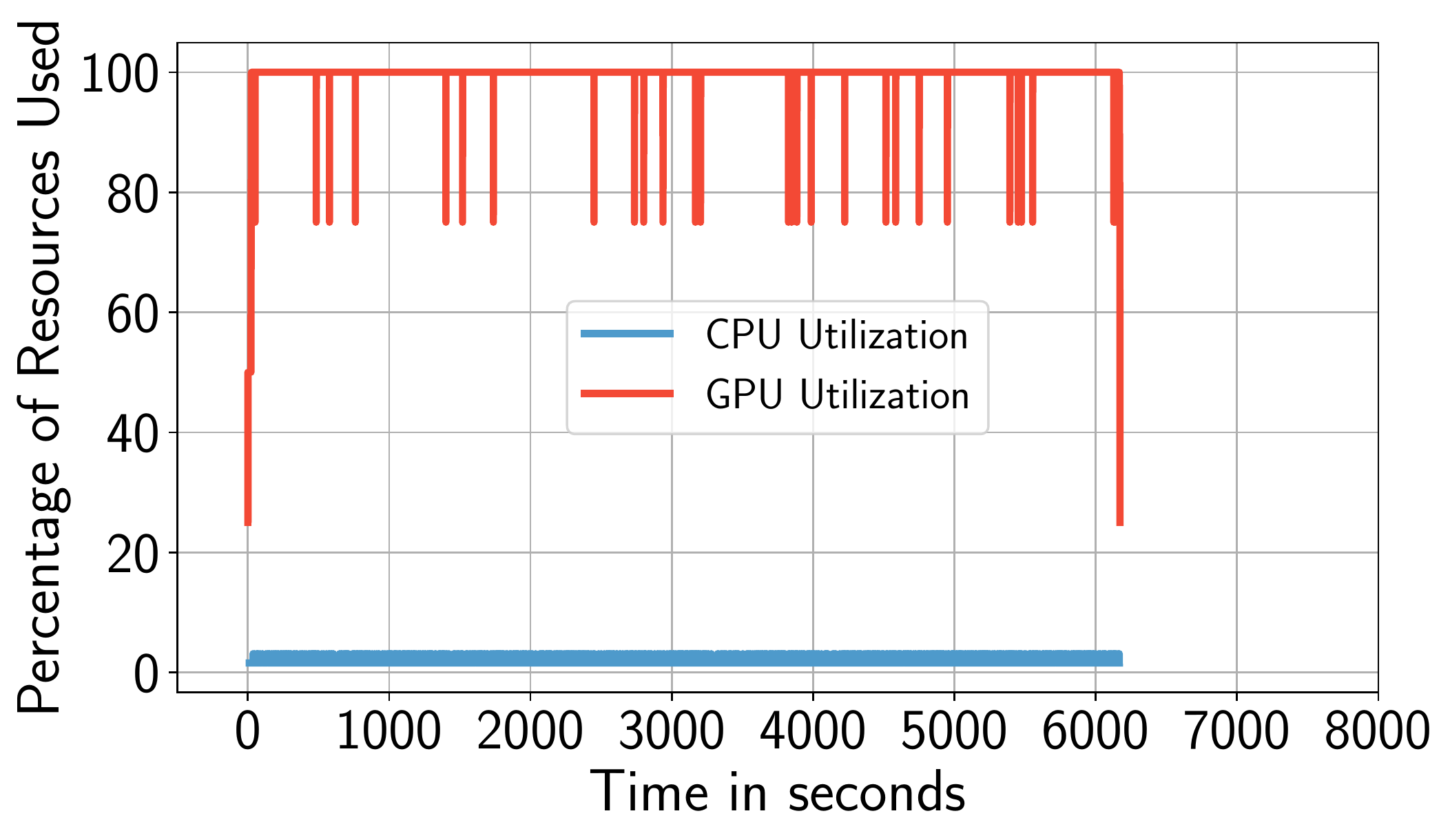}
        \caption{}
        \label{fig:geo_design2autil}
    \end{subfigure}
    \caption{Experiment~2, UC2 Percentage of CPU and GPU utilization for: (a)
    Design~1; (b) Design~2, and (3) Design~2.A.}\label{fig:geo_Utilization}
\end{figure*}

Figs.~\ref{fig:Utilization} and~\ref{fig:geo_Utilization} show the resource
utilization percentage, for all designs, for UC1 and UC2 respectively. CPU
utilization for UC1 with Design~1 (Fig.~\ref{fig:design1util}) closely
approximates the 10.6\% theoretical maximum utilization but GPU utilization
is well below the theoretical 100\%. GPUs are not utilized for almost an hour
at the beginning of the execution and utilization decreases to 80\% some time
after half of the total execution was completed. Our analysis shows that
RADICAL-Pilot's scheduler did not schedule GPU tasks at the start of the
execution even if GPU resources were
available~\cite{paraskevakos2019workflow}.

Fig.~\ref{fig:geo_design1util} shows the resource utilization of UC2 with
Design~1. Average GPU utilization is 97\% and it is reached in about
13.75sec, showing that the issues with GPU execution observed for UC1 were
addressed. Nonetheless, average CPU utilization is only 1\% with large amount
of time spent throttling $T^{UC2}_{2}$ executions. This is explained by the
distribution of $T^{UC2}_{2}$ execution time and the capabilities of the
RTS. The mean execution time of $T^{UC2}_{2}$ is 1.1 seconds and
the task scheduler of the RTS is not able to sustain the throughput required
to use the available cores. While the output of $T^{UC2}_{1}$ tasks
accumulates, $T^{UC2}_{2}$ tasks wait in the scheduler and executor queues of
the RTS.

UC1 does not suffer from the same scheduling limitations of UC2 for Design~1.
The mean execution time of $T^{UC1}_{2}$ is 194 seconds, requiring much less
throughput from the scheduler of the RTS\@. Further, the mean execution time
of $T^{UC1}_{1}$ is 85 seconds instead of the 6 seconds of $T^{UC2}_{1}$.
This produces a much lower output rate for $T^{UC1}_{1}$ than that of
$T^{UC1}_{2}$. In turn, the RTS scheduler has fewer tasks per unit of
time to schedule. Overall, we can conclude that for tasks with less than 1
minute execution time, the overheads of scheduling and setting up the
execution of a task become dominant in the RTS we utilized.

Fig.~\ref{fig:design2util} shows resource utilization for UC1
with Design~2. GPUs are utilized almost immediately as images are becoming
available in the queues between $T^{UC1}_{1}$ and $T^{UC1}_{2}$. This quickly
leads to fully utilized resources. CPU utilization is larger compared
to Design~1, which is expected due to the longer execution times measured.
In addition, two GPUs are used for more than
20,000 seconds compared to other GPUs. This shows that the additional
execution time of that node was only due to the data size and not due to idle
resource time.

Fig.~\ref{fig:geo_design2util} shows resource utilization for UC2 with
Design~2. Compared to Design~1, average GPU utilization improves from 97\% to
99\% but, more relevantly, average CPUs utilization grows to 1.35\% with
almost no throttling. The use of queues, the early binding of data and the
pinning of tasks to nodes, all contribute to reduce the need for throughput
in the RTS scheduler. As a result, this reduces total execution time from
61,310 seconds of Design~1 to the 6,704 seconds of Design~2.

Figs.~\ref{fig:design2autil} and~\ref{fig:geo_design2autil} show that, in
Design~2.A, GPUs are released faster compared to Design~1 and Design~2. This
leads to a GPU utilization above 90\% for both use cases. As already
explained in Experiment~1, this is due to differences in data balancing among
designs. Two design choices are effective for the concurrent execution of
data-driven, compute-intensive and heterogeneous workflows: (1) early binding
of data to node with balanced distribution of image size; and (2) the use of
local filesystems for data sharing among tasks.

Drops in resource utilization are observed in all three designs. In Design~1,
although both CPUs and GPUs were used, in some cases CPU utilization dropped
to 6 cores for UC1 and to 3 GPUs for UC2. Our analysis showed that this
occurred when RADICAL-Pilot scheduled both CPU and GPU tasks, pointing to an
inefficiency in the scheduler implementation. Design~2 and 2.A CPU
utilization drops mostly by one CPU where multiple tasks try to pull from the
queue at the same time. This confirms our conclusions in Experiment~1 about
resource competition between middleware and executing tasks. In all designs,
there is no significant fluctuations in GPU utilization, although there are
more often in Design~1 when CPU and GPUs are used concurrently.

\subsection{Experiment~3: Implementation Overheads}\label{ssec:exp3}

\begin{figure}
    \centering
    \begin{subfigure}{0.45\textwidth}
    \includegraphics[width=\linewidth]{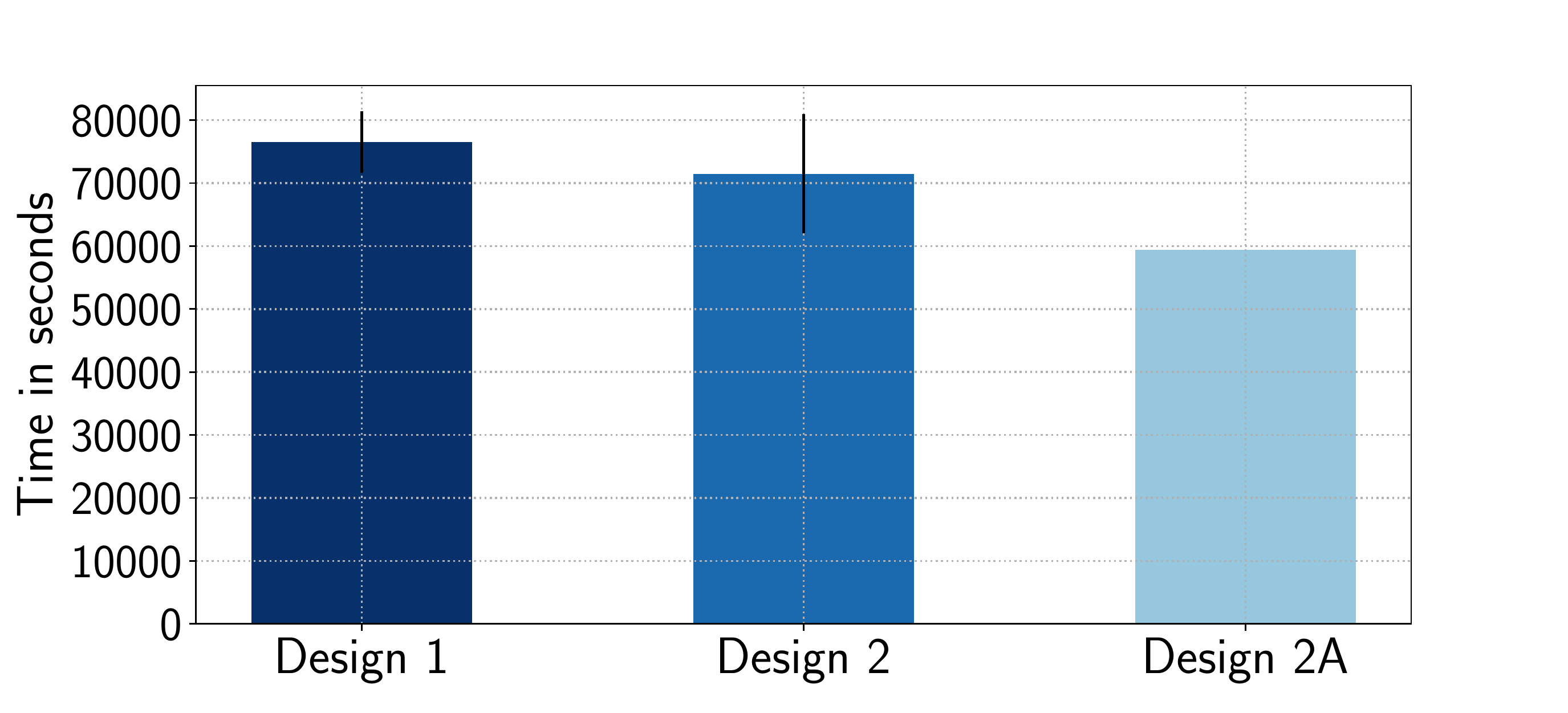}
    \caption{}\label{fig:ttx}
    \end{subfigure}
    ~
    \begin{subfigure}{0.45\textwidth}
    \includegraphics[width=\linewidth]{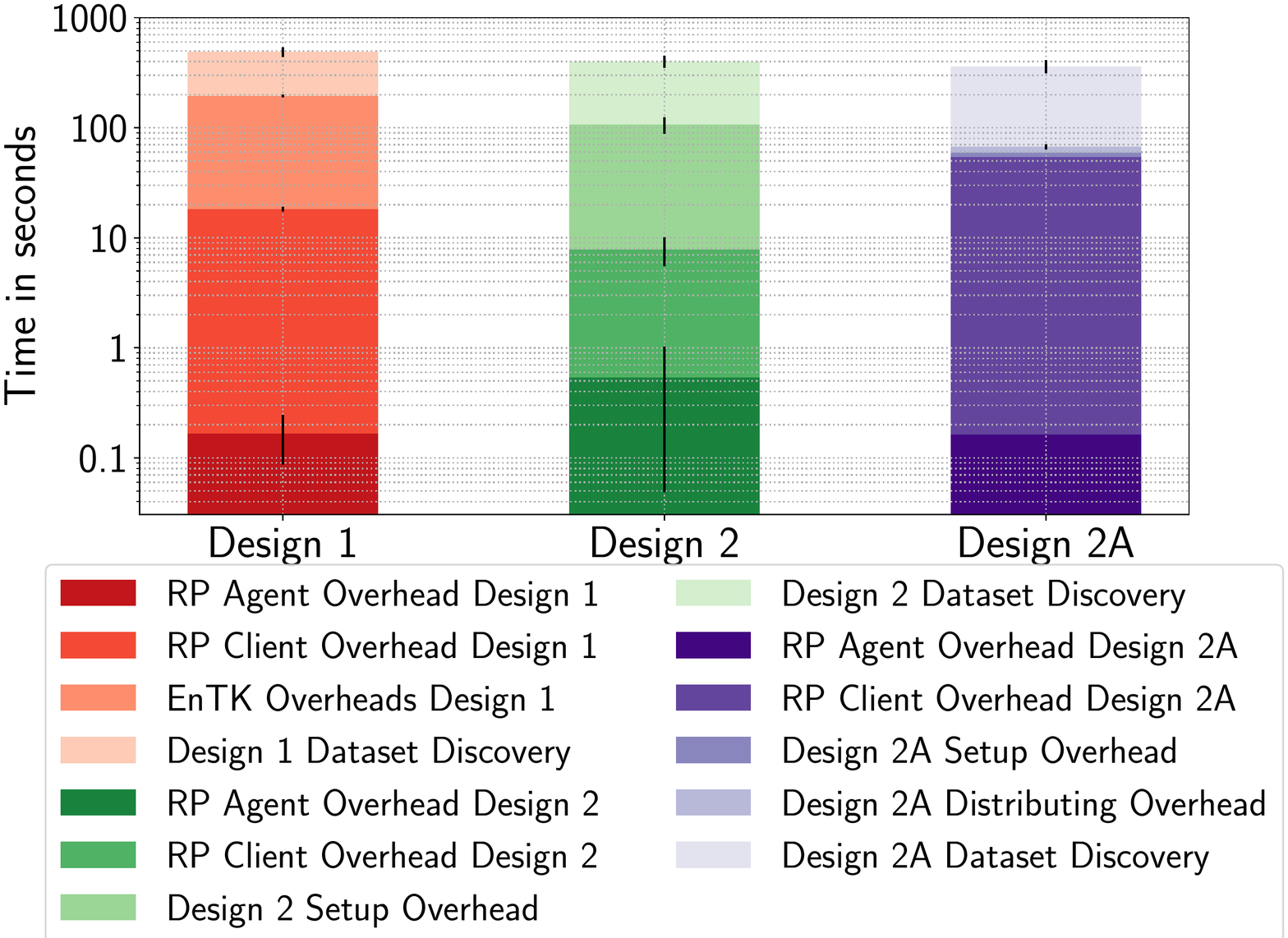}
    \caption{}\label{fig:overheads}
    \end{subfigure}
    \caption{Experiment~3, (a) UC1 total execution time of Design~1, 2 and
    2.A. (b) Overheads of Design~1, 2 and 2.A are at least two orders of
    magnitude less than the total execution
    time.}\label{fig:overall_performance}
\end{figure}

\begin{figure}
    \centering
    \begin{subfigure}{0.45\textwidth}
        \includegraphics[width=\linewidth]{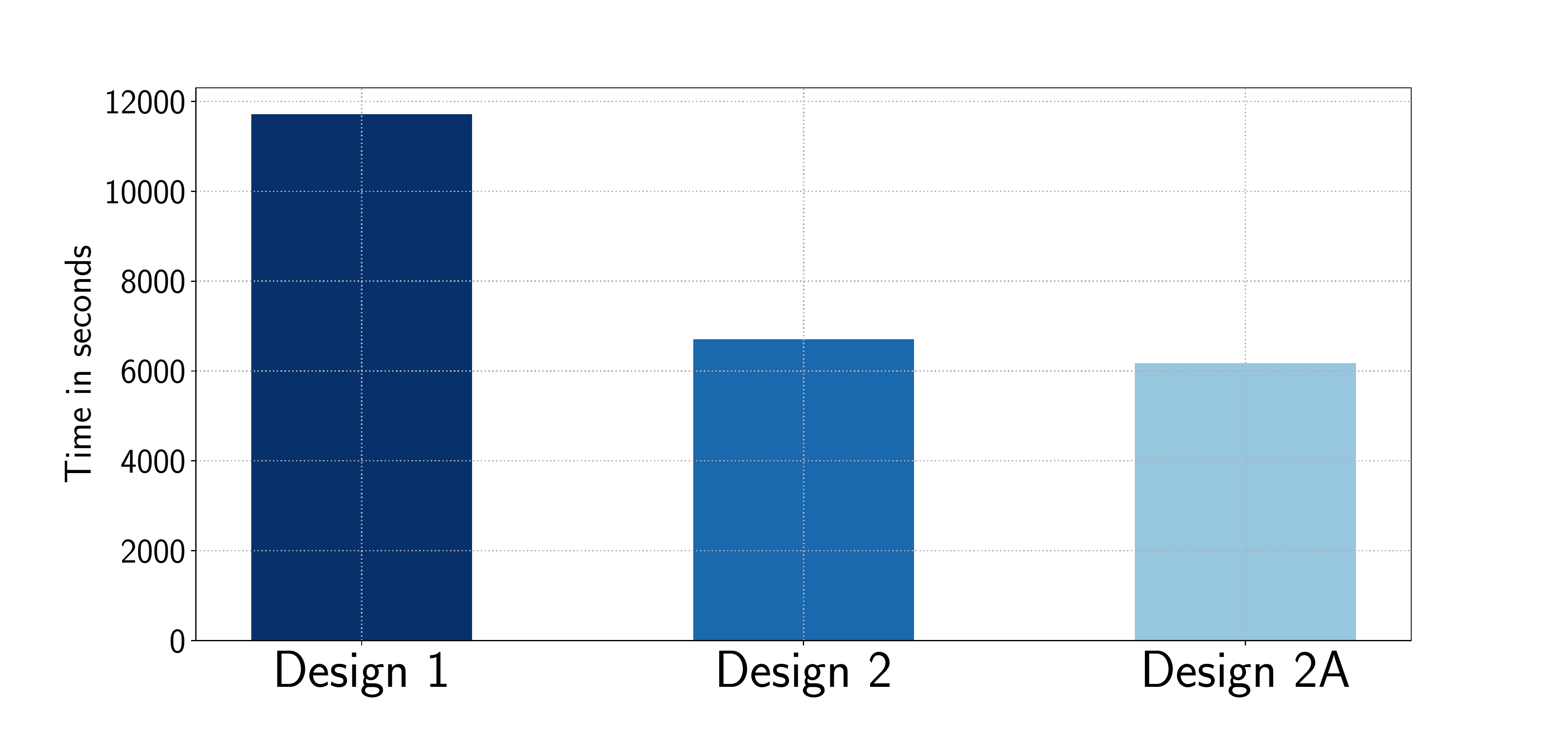}
        \caption{}
        \label{fig:geo_ttx}
    \end{subfigure}
    ~
    \begin{subfigure}{0.45\textwidth}
        \includegraphics[width=\linewidth]{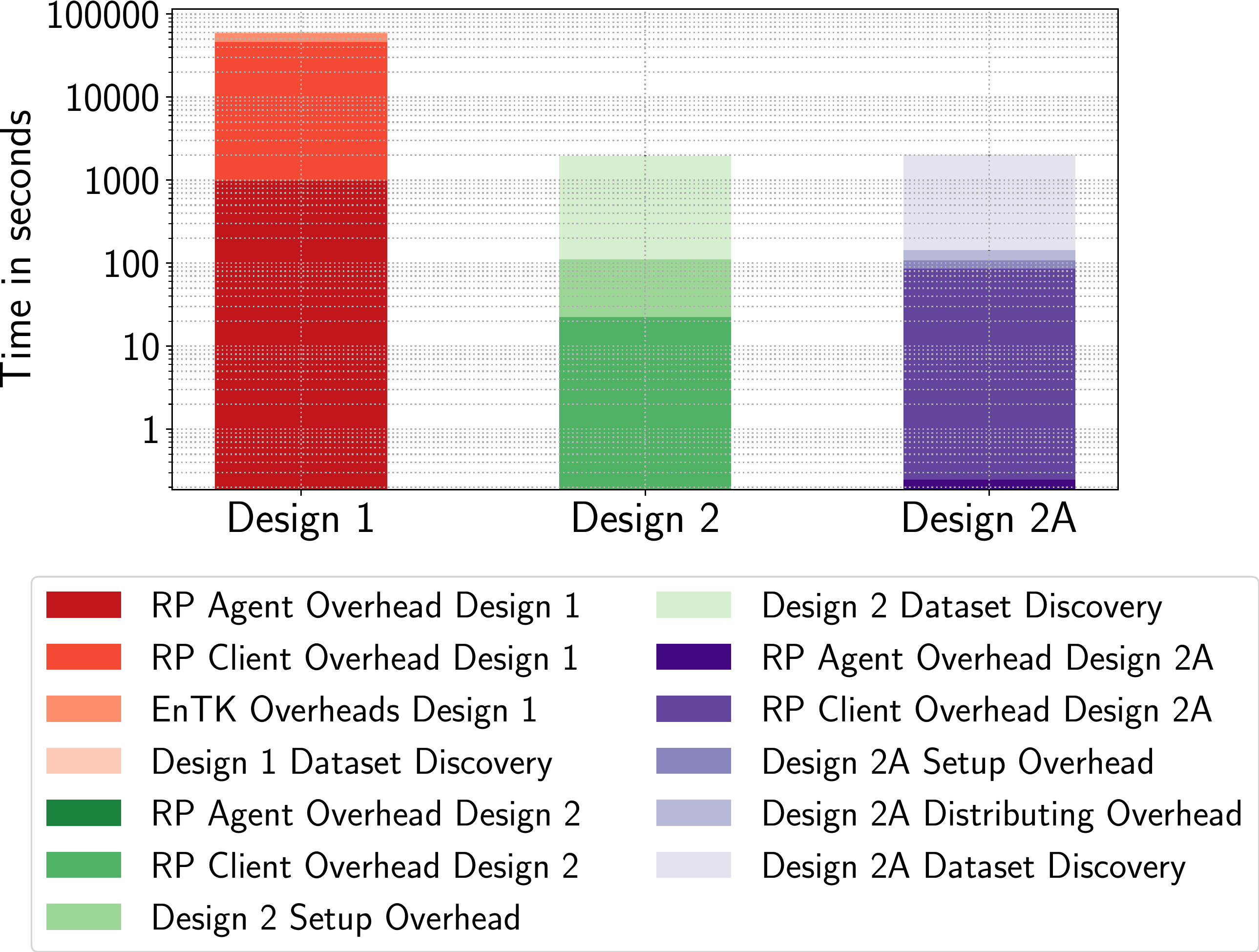}
        \caption{}
        \label{fig:geo_overheads}
    \end{subfigure}
    \caption{Experiment~3, (a) UC2 total execution time of Design~1, 2 and
        2.A. (b) Overheads of Design~1, 2 and 2.A.
    }\label{fig:geo_overall_performance}
\end{figure}

Experiment~3 studies how the total execution time of our use cases workflow
varies across Design~1, 2 and 2.A. Fig.~\ref{fig:ttx} shows that Design~1 and
2 for UC1 have similar total time to execution within error bars, while
Design~2.A is the fastest by a small margin. Fig.~\ref{fig:geo_ttx} shows
that, for UC2, Design~1 total time to execution is around three times longer
than the one of Design~2 and Design~2.A, while Design~2 and Design~2.A have
similar durations. The discussion in \S~\ref{ssec:des1analysis} and
\S~\ref{ssec:des2analysis} explains how these differences relate to the
execution time differences of tasks $T_{1}$ and $T_{2}$, and execution
concurrency.

Figs.~\ref{fig:overheads} and~\ref{fig:geo_overheads} show the overheads of
each design implementation. For UC1, all three designs overheads are at least
two orders of magnitude smaller than the total time to execution. A common
overheads among the three designs is the ``Dataset Discovery Overhead''. This
overhead is the time needed to list the dataset and it is proportional to the
size of the dataset. RADICAL-Pilot has two main components: Agent and Client.
RADICAL-Pilot Agent's overhead is less than a second in all designs while
RADICAL-Pilot Client's overhead is in the order of seconds for all three
designs. The latter overhead is proportional to the number of tasks submitted
simultaneously to RADICAL-Pilot Agent.

EnTK's overhead in Design~1 includes the time to: (1) create the workflow
consisting of independent pipelines; (2) start \entk's components; and (3)
submit the tasks that are ready to be executed to RADICAL-Pilot. This
overhead is proportional to the number of tasks in the first stage of a
pipeline, and the number of pipelines in the workflow. EnTK does not
currently support partial workflow submission, which would allow us to submit
the minimum number of tasks to fully utilize the resources before submitting
the rest.

The dominant overhead of Design~2 is ``Design~2 Setup Overhead''
(Fig.~\ref{fig:overheads}). This overhead includes setting up and starting
queues, and starting and shutting down both tasks $T^{UC1}_{1}$ and 
$T^{UC1}_{2}$ on each compute node. Setting up and starting the queues accounts for
most of the overhead as we use a conservative waiting time to assure that all
the queues are up and ready. This can be optimized further reducing the
impact of this overhead. Design~2.A introduces an overhead called
``Design~2.A Distributing Overhead'' when partitioning and distributing the
dataset over separate nodes. The average time of this overhead for UC1 is
$7.5$ seconds, with a standard deviation of $3.71$ and is proportional to the
dataset and the number of available compute nodes.

Compared to UC1, Fig.~\ref{fig:geo_overheads} shows a different composition
of overheads for UC2. The overheads of Design~1 account for most of the
execution time showed in Fig.~\ref{fig:geo_design1util}. EnTK and RP
Client/Agent overheads are all very large, indicating that RP spends most of
the time scheduling, launching and unscheduling tasks, possibly with very
large I/O overheads due to the high frequency of reading/writing to a shared
file system. EnTK takes a long time waiting for the data required to describe
the full workflow and more time waiting for the tasks to be handled by the
RTS\@. We measured also high latency between the RTS and the
external MongoDB instance used by EnTK and RTS to communicate task
descriptions and state updates. Overall, Design~1 is proven to be unfeasible
for UC2 with the middleware used for our experiments.

Designs~2 and 2.A show much lower overheads than Design~1 for UC2. Note that
the sum of RP and Setup overheads are comparable to those of UC1 but with a
slightly different distribution across overheads components. For UC2, RP
Agent overhead is smaller, possibly due to the improvements made to the RTS
task scheduler. ``Dataset Discovery'' overhead is larger for UC2 compared to
UC1 due to the larger dataset used and the need to form pairs.

In general, Design~2.A offers the best and more stable performance, in terms
of overheads, resource utilization, load balancing and total time to
execution. Although Design~2 has similar overheads, even assuming
minimization of Setup Overhead, it does not guarantee load balancing as done
by Design~2.A. Design~1 involves independent pipelines that
are concurrently executed by the RTS on any available resource,
leading to the described overheads. Based on the results of our analysis,
these overheads could be reduced in both EnTK and RADICAL-Pilot by adopting
early binding of images to each compute node as done in Design~2.A.
Nonetheless, Design~1 would still require executing a task for each image,
imposing bootstrap and tear down overheads for each task.

\section{Conclusions}\label{sec:conclusion}

While Design~1, 2 and 2.A can successfully support the execution of the use
cases described in~\S\ref{sec:ucase}, our experiments show that for the
metrics considered, Design~2.A is the one that offers the better overall
performance. Generalizing this result, use cases that are both data-driven
and compute-intensive benefit from early binding of data to compute nodes so
as to maximize data and compute affinity, and equally balance input data
across nodes. Design~2.A minimizes the overall time to completion of this
type of workflow while maximizing resource utilization.

Our analysis also shows the limits of an approach where pipelines, i.e.,
interdependent compute tasks, are late bound to compute nodes. In designs in
which tasks are independent executables (i.e., programs), the overhead of
bootstrapping a program needs to be minimized, ensuring that each pipeline
processes as much input as possible (in our use case, single and pairs of
images). In presence of large amount of data, late binding implies copying,
replicating or accessing data over network and at runtime. We showed that, in
contemporary HPC infrastructures, this is too costly both for resource
utilization and total time to completion. Even when data are made available
on the network filesystems of the HPC infrastructure, the time spent to
access and/or write those amounts of data at runtime dominates the total time
to completion of the application workflow, vastly reducing the amount of time
computing resources can be used while available.

It should be noted that our insight does not depend on the middleware we used
for our experiments, or on the type of data and computation that our use
cases required. Our insight depends instead on the requirements of the given
tasks and how the capabilities of the available resources satisfy those
requirements. Given the ratio between CPUs and GPUs, the amount of memory per
node and the filesystem performance in our experiments, Design~2.A will
perform better than the other two designs for any use case that requires the
analysis of multi-terabyte dataset with both CPUs and GPUs. Conversely, given
a resource with a sufficiently fast filesystem and a 1:1 ratio between CPUs
and GPUs, based on our analysis, all three designs will perform analogously,
possibly with slightly different overhead distributions.

Infrastructure-wise, the experiments presented in \S\ref{sec:experiments}
show the limits imposed by an imbalance between number of CPU cores and
available memory. Given data-driven computation where multi GB images need
concurrent processing, we were able to use just 10\% of the available cores
due to the amount of RAM required by each image processing. This applies also
to the imbalance between CPUs and GPUs: use cases with heterogeneous tasks
would benefit from a higher GPU/CPU ratio. Finally, filesystem performance
limited the amount of concurrent I/O we could perform from concurrent
processes. This is consistent with the current trend of building HPC
infrastructures with higher GPU density per node, with different types of
dedicated memories and multi-tiered data systems. ORNL Summit or TACC
Frontera are contemporary examples of such a trend.

\S\ref{sec:gpu_imp} and \S\ref{sec:experiments} also show the limits of
optimizing the executable of a task when multiple instances of that
executable have to be executed concurrently. While GPU-SIFT largely improves
on the computing efficiency of preexisting SIFT implementations, the amount
of memory it required to match an image pair always depend on the size of the
images of the pair. This imposes a limit on the number of GPU-SIFT tasks that
can be concurrently executed. This make the use of dedicated accelerators
preferable to the use of general purpose processors, but also shows the
importance of optimizing the design of the middleware that has to execute
those program instances concurrently.

\S\ref{sec:experiments} offers an example of a methodology for experimentally
evaluating the performance of alternative but functionally equivalent
middleware designs that support the execution of data and compute intensive
workflow applications on HPC machines. This methodology is important to drive
the development of middleware in a moment in which application workflows have
become fundamental for many scientific domains~\ref{mcphillips2009scientific}
and academic efforts are multiplying to support such
applications~\ref{workflow-systems-url}.  While qualitative metrics like
usability, security or portability are fundamental, to the best of our
knowledge, we are lacking quantitative ways to compare alternative middleware
designs for specific production infrastructures (i.e., experimental
methodologies). Consistently, our methodology focuses on three quantitative
performance metrics (total time to completion, resource utilization and
middleware overheads) which measure the speed and efficiency with which users
can obtain results and how well resources that have been ``paid'' for can be
utilized.

The results presented open several future lines of research. We will  extend
both \entk{} and RADICAL-Pilot to implement Design~2.A. We will use our
characterization of overheads as a baseline to evaluate our implementations
and further improve the efficiency of our middleware. Further, we will apply
the presented experimental methodology to additional use cases and
infrastructures, measuring the trade offs imposed by other types of task
heterogeneity, including multi-core or multi-GPU tasks that extend beyond a
single compute node. We will explore how the presented methodology applies to
designs in which tasks are not independent programs but, instead, single
functions or methods. In general, we will study how to evaluate the trade
offs between in-memory and filesystem-based computations when use cases
demand the maximization of concurrent execution. This will be particularly
important to evaluate the execution of workflow applications on the upcoming
exascale HPC infrastructures.

Beyond design, methodological and implementation insights, the work for this
paper has already enabled the execution of use cases at unprecedented scale
and speed. The 3,097 images of the Seals use case can be analyzed in
$\approx$20~hours, and the 11,030 image pairs of the Geolocation use case can
be matched in $\approx$5.6~hours, compared to labor-intensive weeks
previously required on non-HPC resources. These are by no means optimal
results. The capabilities we will develop for our middleware based on the
insight gained with the results presented in this paper, will allow for
further improvement of execution performance. In this context, it will be
important to integrate analytical models of optimal execution with the
analysis of executions on actual computing infrastructures.

Data sources, the software used for their analysis and replication guidelines
can be found at~\cite{paraskevakos2019workflow,iceberg-project}.

\section{Acknowledgements and Contributions}
We thank Andre Merzky (Rutgers) and Brad Spitzbart (Stony Brook) for useful
discussions.  This work is funded by NSF EarthCube Award Number 1740572.
Computational resources were provided by NSF XRAC awards TG-MCB090174.
Geospatial support for this work provided by the Polar Geospatial Center under
NSF-OPP awards 1043681 and 1559691. We thank the PSC Bridges PI and
Support Staff for supporting this work through resource reservations.
Aymen Alsaadi, Ioannis Paraskevakos and Matteo Turilli contributed equally to
all section of the paper. Bento Collares Gonçalves contributed to~\S\ref{sec:ucase}.

\bibliographystyle{elsarticle-num}
\bibliography{local}

\end{document}